\documentclass[a4paper]{aa}
\usepackage{graphicx}
\usepackage{txfonts}
\usepackage{color}
\usepackage{natbib}
\usepackage{hyperref}
\usepackage[normalem]{ulem}
\begin{document}

   \title{Abundances of neutron-capture elements\\ in selected solar-type stars}

   \author{Valentina Sheminova  \inst{1},
          Martina Baratella \inst{2},
         Valentina D'Orazi  \inst{3,4}
          }

   \institute{Main Astronomical Observatory, National Academy of Sciences of Ukraine,\\
              Akademika  Zabolotnoho 27,  Kyiv,  03143 Ukraine\\
              \email{shem@mao.kiev.ua}
         \and
              ESO - European Southern Observatory, Alonso de Cordova 3107, Vitacura, Santiago, Chile
          \and
              Department of Physics, University of Rome Tor Vergata, via della ricerca scientifica 1, 00133, Rome, Italy
              \and
              INAF Osservatorio astronomico di Padova, vicolo dell'Osservatorio 5, 35122, Padova, Italy
              }

   \date{Received \hskip 2cm; Accepted}

  \abstract
   {}
{The primary objective of this study is to accurately determine the abundances of Cu, Sr, Y, Zr, Ba, La, and Ce in selected solar-type stars. This will allow us to establish observational abundance--metallicity and abundance--age relations and to  explore the reasons for the excess of Ba compared to other $s$-elements in younger solar-type stars. The chosen $s$-process elements are critical diagnostics for understanding the chemical evolution of our Galaxy.}
{We analysed HARPS spectra with a high resolution ($R$ = 115\,000) and high signal-to-noise ratio (close to 100) of main-sequence solar-type FGK stars with metallicities from $-0.15$ to +0.35 dex and ages from 2 to 14 Gyr using  one-dimensional (1D) local thermodynamic equilibrium (LTE) synthesis and MARCS atmospheric models. In the procedure of fitting synthetic to observed line profiles, the free parameters included abundance and microturbulent and macroturbulent velocity. The macroturbulent velocity can substantially compensate for non-local thermodynamic equilibrium (NLTE) effects in the line core.}
{The resulting elemental abundance [X/H] increases with metallicity and age for solar-type stars.
The ratio of the abundances of $s$-process elements [s/Fe] increases with decreasing metallicity and age, while the [Cu/Fe] ratio increases with both metallicity and age. These observed trends agree well with published observational data and with predictions from  Galactic chemical evolution (GCE) models. A small [Ba/Fe] enhancement of 0.08 ± 0.08 dex has been detected in seven younger stars with an average age of $2.8 \pm 0.6$~Gyr.   Compared to the abundances of other $s$-process elements, [Ba/Fe] is 0.07 and 0.08 dex higher than La and Ce on average, respectively.  Furthermore, we find that the [Ba/Fe] ratio increases with increasing chromospheric activity.
The average [Ba/Fe] for the three most active stars is $0.15 \pm 0.10$~dex higher than that of the other stars. Chromospheric activity, characterised by stronger magnetic fields found in active regions such as pores, spots, plages, and networks, can significantly alter the physical conditions in the formation layers of the Ba lines. Our primary conclusion is that to account for the observed excess of [Ba/Fe] abundance in younger stars, it is essential to use more complex atmospheric models that incorporate magnetic structures.}
   {}


\authorrunning{Sheminova et al.}
   \maketitle
%
%
\section{Introduction}
%
Examining the abundance of chemical elements in stars of different ages helps us to understand how elements are distributed throughout the Galaxy. This study provides insights into the origin and evolution of stellar systems and their interactions with the surrounding environment.

Elements heavier than Fe are produced by neutron-capture processes such as the $s$(slow) and the $r$(rapid) processes.  The foundations for these processes were laid by \citet{1957RvMP...29..547B} and \citet{1955ApJ...121..144C}; for a detailed review, see, e.g. \citet{1994ARA&A..32..153M}, \citet{1999ARA&A..37..239B}, and \citet{2011RvMP...83..157K}. The $s$-process occurs over long time intervals (up to thousands of years fo1r each neutron) with a neutron capture velocity lower than the intermediate $\beta$-decay velocity (one neutron capture per year per nucleus). The $s$-process mainly takes place during the asymptotic branch giant (AGB) phase of the evolution of low-mass stars. On the other hand, the $r$-process occurs on shorter timescales (from fractions of a second to tens of seconds) with a neutron capture velocity of several million to several billion neutron captures per second (faster than the $\beta$-decay).  The $r$-process can occur during supernova explosions, neutron star--neutron star mergers, black hole mergers, and other cosmic events (e.g.  \citealt{2023MNRAS.525.1329L}).

The $s$-process comprises weak, main, and strong components. The weak component primarily produces elements between Cu and Sr and mainly occurs in massive stars ($> 8 M_\odot$) during convective He-core and C-shell burning phases (e.g. \citealt{2010ApJ...710.1557P}). While these elements can also be partially produced by the $r$-process, the $s$-process predominantly causes the abundances observed in the Sun (e.g. \citealt{2022Univ....8..173C}). The main component takes place in low-mass (1.5--4.0~$M_\odot$) AGB stars with near-solar metallicity (e.g. \citealt{2014PASA...31...30K}), as well as in low-metallicity AGB stars with high rotation rates during the thermal-pulse phase (e.g. \citealt{2016MNRAS.456.1803F}). These processes directly influence the chemical evolution. As stars evolve and release their enriched material into the interstellar medium through stellar winds or supernova explosions, the elements synthesised by the $s$-process are distributed throughout the Galaxy. This affects the formation of new stars. Understanding the main $s$-process is crucial for modelling and interpreting the evolution of stars  \citep{1999ARA&A..37..239B}. The strong component of the $s$-process produces heavy elements such as Pb and Bi, and beyond. The strong component is interpreted as the same mechanism that causes the main component, but for the low-metallicity  ([Fe/H]~$< -1.5$), low-mass AGB stars (1--3~$M_\odot$) (\citealt{1998ApJ...497..388G}; \citealt{2011RvMP...83..157K}).
Despite its relatively small contribution to total elemental synthesis, it plays a crucial role in producing elements heavier than  Pb. At low metallicities, Pb becomes the dominant product of low-mass AGB nucleosynthesis, providing a natural explanation for the strong $s$-component \citep{2001ApJ...549..346T}.
The neutron-capture elements are found in three peaks after iron. The
first peak includes the elements at atomic numbers $37\leq Z\leq 40$: Rb, Sr, Y, and Zr. The second peak is formed by the elements with atomic numbers $56 \leq Z \leq 60$: Ba, La, Ce, Pr and Nd, and finally, the elements with $Z=82$, 83 Pb and Bi form the third peak.

It should also be noted that additional contributions to the formation of heavy elements come from the $i$(intermediate) processes \citep{1977ApJ...212..149C} and the $p$(proton) capture process \citep{1994ARA&A..32..153M}. The $i$-process is characterised by neutron densities intermediate between those of the $s$- and $r$-processes.
Rich $i$-process nucleosynthesis can occur during the early AGB phase of low-metallicity low-mass stars \citep{2021A&A...648A.119C}, but different types of stars have been proposed as possible stellar hosts of this process (see \citealt{2021A&A...653A..67B}). The $p$-process can enrich the interstellar medium with  heavy elements such as Mo  and Ru.
It can occur in stellar environments with high proton densities and high temperatures, such as the oxygen-burning envelopes of massive stars just before collapse, during supernovae, and during Type Ia  supernovae. The $p$-process is not as well understood or as thoroughly studied as the $s$-process or the $r$-process, largely due to the complex nuclear physics involved and the challenging conditions under which it occurs \citep{2003PhR...384....1A}.

Studying the abundance of heavy elements in disk stars, in field stars, and in open clusters continues to be of great interest, as evidenced by the large amount of work done on this topic (e.g. \citealt{1993A&A...275..101E}; \citealt{2003MNRAS.340..304R}; \citealt{2006MNRAS.367.1329R}; \citealt{2012Msngr.147...25G}; \citealt{2014A&A...562A..71B}; \citealt{2018PASP..130i4202D}; \citealt{2017A&A...598A..86D};  \citealt{2018A&A...617A.106M}; \citealt{2020A&A...640A..81N}; \citealt{2023MNRAS.525.1329L}; \citealt{2024A&A...683A..75A}).
  Analyses of the abundances in the solar vicinity have shown that about 80\% of Ba (\citealt{2009PASA...26..153S}, \citealt{1999ApJ...521..691T}) and 70\% of La \citep{2006ApJ...647..685W} originate from the $s$-process. As for Cu, it was once thought that most of it was the product of thermonuclear supernovae \citep{1993A&A...272..421M}. However, new calculations have shown that the $s$-process in massive stars can explain the missing Cu abundances \citep{2010ApJ...710.1557P} and  that Cu is mainly formed in hypernovae \citep{2020ApJ...900..179K}.

Recently, \citet{2021A&A...653A..67B} determined $s$-element abundances in young stars of open clusters and found an increase in the [Ba/Fe] ratio, while the [X/Fe] ratios for Sr, Zr, La, and Ce are close to solar values. Similar results have also been obtained for the young open cluster M39 by \citet{2024A&A...683A..75A}.  The Ba excess remains unexplained.
To shed some light on the barium problem, we decided to study solar-type stars in the solar neighbourhood.  Solar-type stars are particularly important in abundance determination studies because the effects of atmospheric parameters are well understood in the Sun, and this allows the values obtained from these stars to be more accurately calibrated using solar measurements. Their surface chemical composition represents the chemical make-up of the molecular cloud from which they were born. They are therefore invaluable astrophysical laboratories, providing unique opportunities to study stellar evolution, planet formation, nucleosynthesis processes, and the discovery of new exoplanets.

This work aims to obtain accurate new abundance measurements of the $s$-elements Cu, Sr, Y, Zr, Ba, La, and Ce in selected solar-type stars in the solar neighbourhood. It also seeks to evaluate how these abundances vary with metallicity, age, and stellar activity. Precise measurements of the $s$-elements are essential for understanding the reasons behind the observed excess of Ba compared to other $s$-elements such as La and Ce in younger stars.

The paper is structured as follows. In Sect.2 we provide information about the selected stars, the spectra, and the lines. Sect.~3 describes the particularities of the stellar abundance measurements. The results and comparison with literature estimates are presented in Sect.~4. Conclusions are given in Sect.~5.

\begin{table}
\centering
 \caption{ESO archive information on stellar spectra and observations.}
\label{tab-obs}
\fontsize{8pt}{9pt}
\selectfont
\begin{tabular}{lccr}
\hline\hline
 Star   &  Date of observation      &      ESO Program ID &     S/N   \\
\hline
HD 189627&  2018-07-15T02:12:52.729  &     100.C-0487(A)   &    150\\
HIP 51987&  2012-05-16T23:52:38.215  &     089.C-0497(A)   &    ~92\\
HD  93849&  2011-03-31T03:48:39.268  &     087.C-0368(A)   &    138\\
HD 158469&  2012-09-14T01:13:47.950  &     089.C-0497(B)   &    139\\
HD 127423&  2008-04-25T06:50:55.613  &     081.C-0148(A)   &    115\\
HD 6790  &  2013-09-27T05:46:55.270  &     091.C-0866(C)   &    ~77\\
HD 102196&  2011-03-31T05:34:50.440  &     087.C-0368(A)   &    139\\
HD 102361&  2011-03-30T04:37:19.474  &     087.C-0368(A)   &    117\\
HD 147873&  2008-09-15T23:52:17.310  &     081.C-0148(B)   &    160\\
HD 38459 &  2008-04-22T23:52:43.466  &     081.C-0148(A)   &    105\\
HD 42936 &  2015-12-23T08:37:18.673  &     096.C-0876(A)   &    ~74\\
HD 221575&  2014-11-19T01:58:53.846  &     192.C-0224(C)   &    ~96\\
HD 128356&  2011-08-13T22:51:09.261  &     087.C-0368(B)   &    129\\
\hline
\end{tabular}
\end{table}
\begin{table*}
\centering
 \caption{Parameters of the selected solar-type stars.}
\label{tab-par}
\small{
\begin{tabular}{llcccccccc}
\hline\hline
Star    & Sp &$T_{\rm eff}$&$\log g$& [Fe/H ] & $t$  &$V\sin i$ & RV     & $R^\prime_{\rm HK}$  &$D$ \\
         &    &   (K)         &     &         &(Gyr)&(km s$^{-1}$)&(km s$^{-1}$)& &(pc)\\
\hline
HD 189627& F7 V      &6210   &4.40  &~~0.07   &~4.0& 5.9& $-14.9$ & $-5.26$& ~70.92  \\
HIP 51987& F0 V      &6158   &5.10  &~~0.27   &~7.2& 2.1& $~-4.9$ & $-5.07$& ~85.25  \\
HD 93849 & G0/1 V    &6153   &4.21  &~~0.08   &~3.5& 3.0& $~~~8.9$& $-5.06$& ~70.27  \\
HD 158469& F8/G2 V   &6105   &4.19  &$-$0.14  &~2.0& 3.1& $~39.4$ & $-5.16$& ~72.15  \\
HD 127423& G0 V      &6020   &4.26  &$-$0.09  &~3.1& 2.5& $-29.7$ & $-4.63$& ~68.31  \\
HD 6790  & G0 V      &6012   &4.40  &$-$0.06  &~3.5& 2.9& $-23.6$ & $-4.92$& 105.93  \\
HD 102196& G2 V      &6012   &3.90  &$-$0.05  &~3.0& 3.6& $~~21.5$& $-5.07$& ~96.34  \\
HD 102361& F8 V      &5978   &4.12  &$-$0.15  &~2.0& 5.0& $-12.9$ & $-4.98$& ~83.61  \\
HD 147873& G1 V      &5972   &3.90  &$-$0.09  &~2.6& 6.5& $~~23.2$& $-4.99$& 104.93  \\
HD 38459 & K1 IV-V   &5233   &4.43  &~~0.06   &~9.0& 1.8& $~~26.2$& $-4.46$& ~35.29  \\
HD 42936 & K0 IV     &5126   &4.44  &~~0.19   &12.0& 1.0& $~~35.9$& $-5.09$& ~46.17  \\
HD 221575& K2 V      &5037   &4.49  &$-$0.11  & 6.0& 1.9& $~~6.46$& $-4.44$&  32.79  \\
HD 128356& K2.5 IV   &4875   &4.58  &~~0.34   &14.0& 1.0& $~~39.7$& $-4.75$& ~26.03  \\
Sun~~~& G2 V         &5777   &4.44  &~~0.00   &~4.6& 1.8& $~~....$& $-4.95$& ~00.00  \\
\hline
\end{tabular}
\tablefoot{ $T_{\rm eff}$ is the effective temperature, $\log g$ is the surface gravity, [Fe/H] is the metallicity, $t$ is  the stellar age, $V\sin i$ is the rotational velocity, RV is the radial velocity, $R^\prime_{\rm HK}$ is the chromospheric activity index, and $D$ is the distance from the Sun.  The stars are listed in descending order of their  $T_{\rm eff}.$}}
\end{table*}

\section{Stellar sample: Spectra and atmospheric parameters}

The observed stellar spectra were taken from the ESO archive (Table~\ref{tab-obs}). They were obtained as part of the  Calan Hertfordshire Extrasolar Planet Search (CHEPS) program, described in detail in \citet{2008A&A...485..571J}, and \citet{2009MNRAS.398..911J}. Observations were made with the High-Accuracy Radial-velocity Planet Searcher (HARPS) spectrometer at a spectral resolution of $R$ = 115\,000. The signal-to-noise ratio (S/N) per resolution element is about 100 on average for our spectra, covering the visible spectrum from 380 to 690 nm.
 The CHEPS program was designed to search for planets, and the sample therefore included stars with a wide range of different metallicities.
  In addition to stellar spectra, we also used observations of the integral solar flux spectrum obtained with HARPS \citep{2013A&A...560A..61M}.  Since this HARPS solar spectrum is only available in two wavelength ranges, 475--531  and 533--566~nm, we also used the Institut für Astrophysik, Göttingen (IAG) solar flux atlas with a resolution of 1\,000\,000 in the wavelength range 405--1065 nm \citep{2016A&A...587A..65R}. This was necessary to extend the list of spectral lines for our abundance analysis. A comparison of the line profiles in the HARPS and IAG spectra, taking into account the spectral resolution, showed good agreement.

\begin{figure*}
   \centering
  \includegraphics[width=0.95\textwidth]{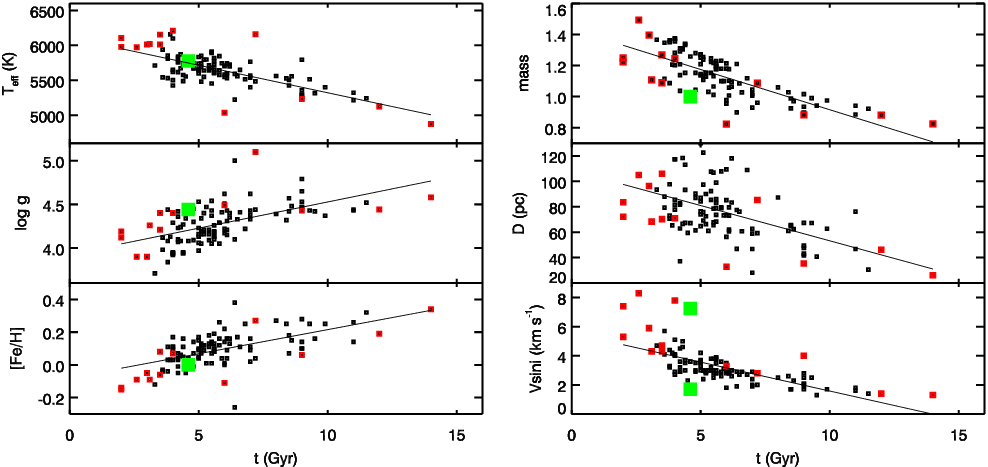}
  \caption {Main parameters of 13 stars (red squares) selected for this analysis from the sample of 107 CHEPS dwarf stars (black squares) vs. stellar age. The Sun is indicated by a large green square. The solid line shows the least-squares fit to the data for all stars.}
 \label{par-t-107}
 \end{figure*}

\begin{figure}[ht]
 \centering
  \includegraphics[width=0.35\textwidth]{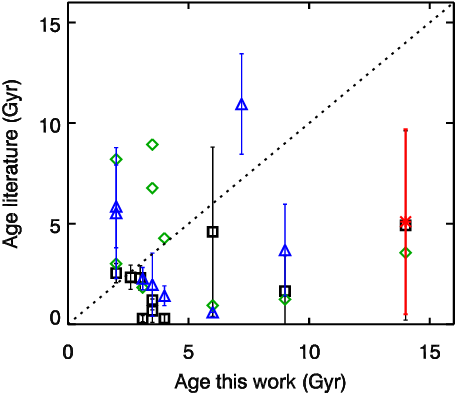}
   \caption {Comparison between the ages used in the present work, derived by  \citet{2019A&A...621A.112P}, and the ages derived by  \citet{2021A&A...649A.111A}, \citet{2018AJ....155..111L},  \citet{2021A&A...646A..77G},
   \citet{2020ApJ...898...27S} noted by  red, blue, black, and green symbols, respectively.}
\label{age-all}
 \end{figure}

Stellar parameters for solar-type stars with ages exceeding 1~Gyr can be determined with a remarkable level of accuracy. For consistency, we selected stars from the CHEPS program considering the atmospheric stellar parameters, the chromospheric index ($R^\prime_{\rm HK}$), and the ages ($t$),  and the distance ($D$) from the Sun using the uniformly determined values from \citet{2017MNRAS.468.4151I}, \citet{2008A&A...485..571J}, and \citet{2019A&A...621A.112P}, respectively. On the other hand, the rotational velocity ($V\sin i$) and the radial velocity (RV) have been measured in the works of \citet{2019KPCB...35..129S, 2022KPCB...38...83S}.
 Our stellar sample comprises 13 stars with effective temperatures ($T_{\rm eff}$) from 4800 to 6200 K, surface gravities ($\log g$) from 3.9 to 5 dex, and metallicities ([Fe/H]) from $-0.15$ to 0.35 dex. Figure~\ref{par-t-107} shows the main properties of our targets (red squares) stars and the full sample from \citet{2017MNRAS.468.4151I} (black squares) compared to the Sun (large green square). The main parameters are reported in  Table~\ref{tab-par}, where the spectral type is taken from SIMBAD data \footnote{http://cdsportal.u-strasbg.fr}.
We have used the estimates of stellar ages obtained by \citet{2019A&A...621A.112P} from the luminosities and evolutionary tracks for the sample of 107 stars. They have large errors, especially for stars with masses lower than one solar mass. For some stars, no proper solutions were found, that is, their calculated ages exceed the Hubble time ($>14$~Gyr). For example, an estimate of 15.5~Gyr was obtained for our low-mass star HD 128356. We therefore imposed a limit age of 14 Gyr for this star. The estimated ages given in the literature for some of our stars are also subject to large uncertainties (see  Fig.~\ref{age-all}). The discrepancies in the stellar age estimates used by us and obtained from various literature sources increase with age. The oldest star in our sample, HD 128356, has an uncertainty in age of 4.6 Gyr according to \citet{2021A&A...649A.111A}.
\begin{table}
\centering
 \caption{List of lines used in this analysis.}
\label{tab-line}
\small{
\begin{tabular}{clcrcccc}
\hline\hline
$\lambda$ &  El &  $E_{\rm exp}$ &$\log gf$&  $ A_\odot$ &  $\log \tau_5$& HFS\\
  (nm)    &     & (eV)           &         &             &               &    \\
\hline
 444.5471 & Fe I &0.087&$-$5.441  &    7.53 &$ -2.09$ & n \\
 499.4129 & Fe I &0.915&$-$3.080  &    7.51 &$ -4.03$ & n \\
 524.2491 & Fe I &3.634&$-$0.967  &    7.56 &$ -2.86$ & n \\
 537.9574 & Fe I &3.695&$-$1.514  &    7.57 &$ -2.15$ & n \\
 608.2710 & Fe I &2.223&$-$3.573  &    7 49 &$ -1.70$ & n \\
 615.1618 & Fe I &2.176&$-$3.299  &    7 53 &$ -2.08$ & n \\
 625.2555 & Fe I &2.404&$-$1.687  &    7.55 &$ -3.78$ & n \\
 510.5541 & Cu I &1.389&$-$1.520  &    4.11 &$ -2.44$ & y \\
 522.0070 & Cu I &3.817&$-$0.671  &    4.13 &$ -1.10$ & y \\
 578.2132 & Cu I &1.642&$-$1.720  &    4.15 &$ -1.64$ & y \\
 421.5524 & Sr II&0.000&$-$1.173  &    2.80 &$ -4.90$ & y \\
 460.7331 & Sr I &0.000&$ $0.283  &    2.70 &$ -2.09$ & n \\
 439.8008 & Y II &0.129&$-$1.000  &    2.21 &$ -2.60$ & n \\
 488.3682 & Y II &1.083&$ $0.070  &    2.24 &$ -2.69$ & n \\
 508.7418 & Y II &1.083&$-$0.170  &    2.15 &$ -2.24$ & n \\
 405.0330 & Zr II&0.713&$-$1.060  &    2.52 &$ -1.41$ & n \\
 420.8980 & Zr II&0.713&$-$0.510  &    2.59 &$ -2.16$ & n \\
 511.2270 & Zr II&1.665&$-$0.850  &    2.56 &$ -1.10$ & n \\
 455.4033 & Ba II&0.000&$ $0.140  &    2.27 &$ -4.75$ & y \\
 585.3675 & Ba II&0.604&$-$1.010  &    2.27 &$ -3.04$ & y \\
 649.6898 & Ba II&0.604&$-$0.406  &    2.37 &$ -4.04$ & y \\
 398.8520 & La II&0.403&$ $0.210  &    1.11 &$ -1.26$ & y \\
 399.5750 & La II&0.173&$-$0.010  &    1.09 &$ -1.35$ & y \\
 408.6720 & La II&0.000&$-$0.070  &    1.13 &$ -2.02$ & y \\
 399.9240 & Ce II&0.295&$ $0.060  &    1.50 &$ -1.47$ & n \\
 407.3480 & Ce II&0.478&$ $0.210  &    1.56 &$ -1.47$ & n \\
 456.2360 & Ce II&0.478&$ $0.210  &    1.60 &$ -1.59$ & n \\
 527.4230 & Ce II&1.044&$ $0.130  &    1.53 &$ -1.24$ & n \\
 \hline
\end{tabular}
\tablefoot{$\lambda$ is  the wavelength, $E_{\rm exp}$ is the excitation potential, and $\log gf$ is the oscillator strength. The data of the solar abundance $A_\odot$ and the effective depth of the line core formation $\log \tau_5$ were derived from this analysis. The last column indicates whether the line has a hyperfine structure.}
}
\end{table}

\section{Abundance determination}

To reliably determine the abundance of a given species through spectral synthesis, it is crucial to use as many spectral lines of the $s$-elements as possible within the wavelength range. We selected appropriate lines for analysis, ensuring they were strong enough to be measured accurately, even in the hottest and most metal-poor stars in our sample.
 For example, Zr is present in the solar photosphere in the form of neutral and ionised species. However, the lines of Zr I are too weak to be measured in all our stars. Well-measured lines in the solar spectrum of other elements in the studied wavelength range, such as Y and La, are also very weak or blended in our targets. Our list includes all lines that were selected by \citet{2021A&A...653A..67B}. We added to the list some lines with blends in the distant wings that  are clearly visible in the spectra of cool stars and can be taken into account through the spectral synthesis.

Additionally, seven Fe I lines with accurate oscillator strengths from \citet{2006JPCRD..35.1669F} were added to our list.
We decided to re-derive the iron abundance [Fe/H] in the exact same way as the $s$-process elements so that the [s/Fe] values were determined in a uniform manner. The complete line list is given in Table~\ref{tab-line}. Atomic data of the selected lines and the data necessary to account for the hyperfine structure and isotropic composition were taken from the VALD3 database \citep{2015PhyS...90e4005R}. Atomic data of nearby lines and blends for the synthesis of spectral regions were taken from the database of atomic spectral lines \citep{1995KurCD..23.....K}. Molecular lines are not considered in this analysis, since they are almost insignificant in the selected lines, except for the very strong 421.55 nm Sr II line, in which the CN molecular lines are more pronounced.

Among the $s$-elements, we chose  Cu, Sr, Y, Zr, Ba, La, and Ce  based on  previous analyses (e.g. \citealt{2021A&A...653A..67B}). The stellar abundances were determined differentially with respect to the solar values on a line-by-line basis.  Using the same spectral line of a given element X, we computed its stellar abundance $A_\star$ and its solar abundance $A_{\odot}$, where $A$ is in the usual form $A= 12+\log$(X/H). Then, we obtained the relative stellar abundance [X/H] = $\log (A_\star$(X)$/ A_\odot$(X)), and the metallicity of the star as [Fe/H] = $\log (A_\star$(Fe)$/ A_\odot$(Fe)). In this way, any uncertainty with atomic data (e.g. the oscillator strengths) was minimised. In addition, we obtained the ratio of the abundance of the X element to the abundance of the reference element Fe, namely [X/Fe]~=~[X/H]~$-$~[Fe/H].
The abundances were derived through the spectral synthesis using the SPANSAT LTE code developed by \citet{1988ITF....87P...3G} and one-dimensional (1D) models of stellar atmospheres obtained by interpolation from the MARCS database of \citet{2008A&A...486..951G}.  Models of the atmospheres of the Sun and stars were recalculated for the chemical composition from  \citet{2019arXiv191200844L}.
A simple model of isotropic depth-independent micro- and macroturbulence was used for the line broadening. The radial-tangential (RT) model, especially for the macroturbulence velocity, is better for most stars in the Hertzsprung–Russell diagram.   For solar-type stars, the difference between using the isotropic and RT models in a line synthesis is small \citep{2017KPCB...33..217S}. We have tested these models and ensured that the difference does not significantly affect the derived abundances. The model atmosphere was re-computed each time a new $s$-process abundance was found, and the chemical composition was updated.

\begin{table*}
\centering
 \caption{Solar abundances $A_\odot$ of Ba derived in this work and by other authors who used 1D NLTE and three-dimensional (3D) NTLE synthesis.}
 \label{tab-nlte}
 {\small
\begin{tabular}{lccccc}
\hline\hline
Ba II lines   &    This work  &  G20$^{\rm 1}$ & G15$^{\rm 2}$    & G20$^{\rm 1}$& G15$^{\rm 2}$    \\
    (nm) &    (1D LTE)  & (1D NTLE)&  (1D NTLE)&    (3D NTLE) & (3D NTLE) \\
\hline\hline
455.4 &    2.27      &   2.16   &  2.13     &    2.29      & 2.22     \\
585.3 &    2.27      &   2.13   &  2.12     &    2.27      & 2.31     \\
649.6 &    2.37      &   2.06   &  2.04     &    2.27      & 2.22     \\
\hline
\end{tabular}

$^{\rm 1}$ \citet{2020A&A...634A..55G},
$^{\rm 2}$ \citet{2015A&A...573A..27G}
}
\end{table*}
\begin{table*}
\centering
 \caption{Solar abundances $A_\odot$ of the analysed elements derived by us and other authors.}
\label{tab-a-sun}
{\small
\begin{tabular}{lcccccc}
\hline\hline
  El  & This work   & G15 $^{\rm 1}$& L19$^{\rm 2}$ & B21$^{\rm 3}$  & DM19$^{\rm 4}$ & M15$^{\rm 5}$  \\
\hline\hline
  Fe & 7.53$\pm$0.03&  7.47$\pm$0.04&  7.52$\pm$0.05&  7.45$\pm$0.01 &    $....$& 7.57$\pm$ 0.00\\
  Cu & 4.13$\pm$0.02&  4.18$\pm$0.05&  4.21$\pm$0.03&  4.21$\pm$0.12 &    4.10& $....$         \\
  Sr & 2.75$\pm$0.07&  2.83$\pm$0.06&  2.92$\pm$0.05&  2.85$\pm$0.03 &    2.78&  $....$       \\
   Y & 2.20$\pm$0.05&  2.21$\pm$0.05&  2.20$\pm$0.05&  2.20$\pm$0.03 &    2.21& 2.15$\pm$ 0.17\\
  Zr & 2.56$\pm$0.04&  2.59$\pm$0.04&  2.59$\pm$0.06&  2.57$\pm$0.01 &    2.65& 2.79$\pm$ 0.19\\
  Ba & 2.30$\pm$0.06&  2.25$\pm$0.07&  2.19$\pm$0.07&  2.31$\pm$0.09 &    2.25& 2.17$\pm$ 0.04\\
  La & 1.11$\pm$0.02&  1.11$\pm$0.04&  1.14$\pm$0.03&  1.08$\pm$0.03 &    $....$& 1.24$\pm$ 0.02\\
  Ce & 1.55$\pm$0.04&  1.58$\pm$0.04&  1.61$\pm$0.06&  1.60$\pm$0.05 &    1.60& 1.70$\pm$ 0.11\\
\hline
\end{tabular}

$^{\rm 1}$ \citet{2015A&A...573A..27G},
$^{\rm 2}$ \citet{2019arXiv191200844L},
$^{\rm 3}$ \citet{2021A&A...653A..67B},

$^{\rm 4}$ \citet{2019A&A...624A..78D},
$^{\rm 5}$ \citet{2015MNRAS.446.3651M}
}
\end{table*}

The van der Waals damping constants were calculated using the Anstee-Barklem-O'Mara method. The required damping parameters $\sigma$ and $\alpha$ for Fe lines were taken from the tables of  \citet{2005A&A...435..373B} and \citet{2000A&AS..142..467B}, and from the VALD3 database for the lines of other elements. The hyperfine and isotopic structures of lines of heavy elements (see Table~\ref{tab-line}) were taken into account in the line synthesis and the computations of effective line formation depths. The algorithm for computing effective depths using depression functions was described in  \citet{2021A&A...653A..67B}.

The synthetic profile was fitted to the observed profile by finding the lowest value of $\chi^2= \sum^N_{k=1} ((R_{k,\rm ods}- R_{k,\rm syn})^2/R_{k,\rm syn})$. Here, the flux profile $R=F/F_c$,  $N$ is the number of points of the profile fitting. To obtain the best fit, we used three free parameters, namely the abundance $A$, the microturbulent velocity $\xi$, and the macroturbulent velocity $\zeta$. Each of these parameters has a different effect on the line profile. The microturbulence has a stronger effect on the half-width of the line profile, macroturbulence predominantly affects the core and far wings, and abundance acts on the central part of the profile.  The profile-fitting procedure was performed manually with visual inspection through several iterations.
\begin{table*}
\centering
 \caption{Abundances [X/H]  and  the relative  abundances [X/Fe] in selected solar-type stars.}
\label{tab-x-h}
\fontsize{8pt}{9pt}
\selectfont
\begin{tabular}{lrrrrrrrr}
\hline\hline
 Star    &        [Cu/H]   &       [Sr/H]     &     [Y/H]      &  [Zr/H]       &       [Ba/H]       &  [La/H]       &     [Ce/H]      &      [Fe/H]    \\
\hline
HD 189627&   $ 0.15\pm0.02$ &  $ 0.10\pm0.02$ &  $ 0.12\pm0.05$&  $ 0.17\pm0.03$&   $ 0.05\pm0.02$  & $ 0.14\pm0.05$&   $ 0.14\pm0.04$&  $~0.08\pm0.03$\\
HIP 51987&   $ 0.51\pm0.07$ &  $ 0.36\pm0.01$ &  $ 0.25\pm0.04$&  $ 0.30\pm0.02$&   $ 0.32\pm0.03$  & $ 0.34\pm0.02$&   $ 0.36\pm0.04$&  $~0.36\pm0.04$\\
HD  93849&   $ 0.26\pm0.03$ &  $ 0.17\pm0.00$ &  $ 0.01\pm0.04$&  $ 0.10\pm0.05$&   $ 0.15\pm0.07$  & $ 0.07\pm0.01$&   $ 0.04\pm0.03$&  $~0.12\pm0.02$\\
HD 158469&   $-0.10\pm0.01$ &  $-0.12\pm0.01$ &  $-0.14\pm0.04$&  $-0.07\pm0.01$&   $-0.07\pm0.02$  & $-0.04\pm0.02$&   $-0.09\pm0.03$&  $~0.08\pm0.03$\\
HD 127423&   $-0.09\pm0.01$ &  $-0.01\pm0.05$ &  $-0.05\pm0.03$&  $-0.09\pm0.04$&   $ 0.11\pm0.03$  & $-0.09\pm0.00$&   $-0.09\pm0.06$&  $-0.10\pm0.04$\\
HD   6790&   $-0.08\pm0.02$ &  $ 0.01\pm0.03$ &  $-0.03\pm0.03$&  $-0.11\pm0.03$&   $ 0.12\pm0.04$  & $-0.05\pm0.02$&   $-0.05\pm0.06$&  $~0.02\pm0.08$\\
HD 102196&   $ 0.03\pm0.01$ &  $ 0.06\pm0.01$ &  $-0.05\pm0.02$&  $ 0.03\pm0.04$&   $ 0.08\pm0.03$  & $ 0.00\pm0.03$&   $-0.01\pm0.04$&  $-0.03\pm0.03$\\
HD 102361&   $-0.04\pm0.03$ &  $-0.10\pm0.03$ &  $-0.15\pm0.04$&  $-0.12\pm0.02$&   $-0.18\pm0.05$  & $-0.18\pm0.04$&   $-0.18\pm0.06$&  $-0.16\pm0.02$\\
HD 147873&   $ 0.03\pm0.03$ &  $ 0.05\pm0.07$ &  $ 0.08\pm0.04$&  $ 0.03\pm0.03$&   $-0.01\pm0.05$  & $ 0.03\pm0.04$&   $ 0.01\pm0.03$&  $-0.12\pm0.04$\\
HD  38459&   $ 0.06\pm0.08$ &  $ 0.21\pm0.01$ &  $ 0.05\pm0.02$&  $ 0.10\pm0.01$&   $ 0.10\pm0.06$  & $ 0.05\pm0.02$&   $ 0.05\pm0.01$&  $~0.07\pm0.02$\\
HD  42935&   $ 0.21\pm0.05$ &  $ 0.22\pm0.05$ &  $-0.01\pm0.01$&  $ 0.08\pm0.01$&   $ 0.06\pm0.04$  & $ 0.06\pm0.05$&   $ 0.10\pm0.06$&  $~0.16\pm0.03$\\
HD 221575&   $-0.13\pm0.04$ &  $ 0.05\pm0.04$ &  $-0.05\pm0.02$&  $ 0.11\pm0.02$&   $ 0.09\pm0.04$  & $ 0.02\pm0.02$&   $ 0.10\pm0.03$&  $-0.12\pm0.04$\\
HD 128358&   $ 0.44\pm0.07$ &  $ 0.23\pm0.05$ &  $ 0.15\pm0.01$&  $ 0.22\pm0.04$&   $ 0.18\pm0.04$  & $ 0.14\pm0.03$&   $ 0.20\pm0.03$&  $~0.28\pm0.04$\\
\hline\hline
 Star     &       [Cu/Fe]   &     [Sr/Fe]   &     [Y/Fe]       &    [Zr/Fe]     &       [Ba/Fe]   &     [La/Fe]   &     [Ce/Fe]        \\
\hline
HD 189627&   $ 0.07\pm0.04$ & $ 0.02\pm0.04$&   $ 0.04\pm0.06$ & $ 0.09\pm0.04$&   $-0.03\pm0.04$  & $ 0.06\pm0.06$ &   $ 0.06\pm0.05$ \\
HIP 51987&   $ 0.15\pm0.08$ & $ 0.00\pm0.04$&   $-0.11\pm0.06$ & $-0.06\pm0.04$&   $-0.04\pm0.05$  & $-0.02\pm0.04$ &   $ 0.00\pm0.06$ \\
HD  93849&   $ 0.14\pm0.04$ & $ 0.05\pm0.02$&   $-0.11\pm0.04$ & $-0.02\pm0.05$&   $ 0.03\pm0.07$  & $-0.05\pm0.02$ &   $-0.08\pm0.04$ \\
HD 158469&   $-0.02\pm0.03$ & $-0.04\pm0.03$&   $-0.06\pm0.05$ & $ 0.01\pm0.03$&   $ 0.01\pm0.04$  & $ 0.04\pm0.04$ &   $-0.01\pm0.04$ \\
HD 127423&   $ 0.01\pm0.04$ & $ 0.09\pm0.06$&   $ 0.05\pm0.05$ & $ 0.01\pm0.06$&   $ 0.21\pm0.05$  & $ 0.01\pm0.04$ &   $ 0.01\pm0.07$ \\
HD   6790&   $-0.10\pm0.08$ & $-0.01\pm0.09$&   $-0.05\pm0.09$ & $-0.13\pm0.09$&   $ 0.10\pm0.09$  & $-0.07\pm0.08$ &   $-0.07\pm0.10$ \\
HD 102196&   $ 0.06\pm0.03$ & $ 0.09\pm0.03$&   $-0.02\pm0.04$ & $ 0.06\pm0.05$&   $ 0.11\pm0.04$  & $ 0.03\pm0.04$ &   $ 0.02\pm0.05$ \\
HD 102361&   $ 0.12\pm0.04$ & $ 0.06\pm0.04$&   $ 0.01\pm0.04$ & $ 0.04\pm0.03$&   $-0.02\pm0.05$  & $-0.02\pm0.04$ &   $-0.02\pm0.06$ \\
HD 147873&   $ 0.15\pm0.05$ & $ 0.17\pm0.08$&   $ 0.20\pm0.06$ & $ 0.15\pm0.05$&   $ 0.11\pm0.06$  & $ 0.15\pm0.06$ &   $ 0.13\pm0.05$ \\
HD  38459&   $-0.01\pm0.08$ & $ 0.14\pm0.02$&   $-0.02\pm0.03$ & $ 0.03\pm0.02$&   $ 0.03\pm0.06$  & $-0.02\pm0.03$ &   $-0.02\pm0.02$ \\
HD  42935&   $ 0.05\pm0.06$ & $ 0.06\pm0.06$&   $-0.17\pm0.03$ & $-0.08\pm0.03$&   $-0.10\pm0.05$  & $-0.10\pm0.06$ &   $-0.06\pm0.07$ \\
HD 221575&   $-0.01\pm0.06$ & $ 0.17\pm0.06$&   $ 0.07\pm0.04$ & $ 0.23\pm0.04$&   $ 0.21\pm0.06$  & $ 0.14\pm0.04$ &   $ 0.22\pm0.05$ \\
HD 128358&   $ 0.16\pm0.08$ & $-0.05\pm0.06$&   $-0.13\pm0.04$ & $-0.06\pm0.06$&   $-0.10\pm0.06$  & $-0.14\pm0.05$ &   $-0.08\pm0.05$ \\
\hline
\end{tabular}
\end{table*}

The internal uncertainty of the resulting $A$ abundance is determined by many factors:
\begin{enumerate}
\item  One-dimensional (1D) atmospheric models produce a symmetric profile instead of an asymmetric observed profile, which can affect the abundance derived from line profiles. For example, according to \citet{2020A&A...634A..55G}, the difference between solar abundances derived using 3D and 1D models in the LTE approximation is 0.09, 0.19, and 0.19 dex for the Ba II lines 455.40, 585.36, and 649.69 nm, respectively.
\item The distortion of the shape of the observed profiles at a signal-to-noise ratio of about 100 is on average insignificant for our stars. After smoothing,  noises are visible in the line profiles (see Figs.~\ref{prof-lines}, \ref{prof-lines1}). {In the spectra of some stars the noise is very low, while in the spectra of others it is slightly higher.}  The possible errors do not exceed an average of $\pm 0.03$ dex in this analysis.
\item Uncertainties in the determination of the continuum level in observational spectra can introduce errors into the line profile-fitting procedure. By refining the local continuum for each line, we were able to reduce these uncertainties so that they became almost insignificant.
\item Errors due to blending with nearby features are generally small because of the spectral synthesis technique, and they amount to $-$0.01, $-$0.03, and $-$0.04 dex for the Ba II solar lines at 455.40, 585.36, and 649.69 nm, respectively (\citealt{2020A&A...634A..55G}).
\item The accuracy of the abundance determination is affected by the choice of the initial values of the free parameters. We used the initial values of the free parameters obtained earlier in the Fourier analysis for these stars \citep{2017KPCB...33..217S} in order to reduce possible errors.
\item A potential source of abundance errors are molecular bands, which mostly populate the spectra of the cool stars. To reduce these effects, we directly selected lines from the spectrum of the coolest star in our sample, which reduced the impact on the abundance determination for the other stars.
\item The uncertainties due to the LTE assumption are relatively small for most of our lines, and they do not exceed the acceptable error limits of this analysis. However, in the case of the Sr, Y, and Ba lines, they can be significant. For example, the correction for NLTE effects for the Sr 460.7 nm solar line is 0.10 dex \citep{2012A&A...546A..90B}, and for the Ba II solar lines 455.40, 585.36, and 649.69 nm, it is $-0.05$, $-0.11$,  and $-0.19$ dex \citep{2020A&A...634A..55G}. For the Y II lines, the corrections do not exceed 0.12 dex for FGK type stars with a metallicity close to that of the Sun \citep{2023ApJ...957...10A}.
\end{enumerate}

In this analysis, we did not apply NLTE corrections to the derived abundance values.  Instead, potential errors were compensated for by the use of macroturbulence in the profile fitting procedure. The macroturbulence affects the profile core in a similar way to NLTE effects, making the line core either deeper or shallower. By varying $\zeta$, it is possible to achieve the best agreement between the synthetic and observed profiles in the core of the line. As a result, the $\zeta$ value may be lower or higher than the actual value, while the abundance remains the same. In principle, this is not an ideal compensation for NLTE. As $\zeta$ decreases, the line wings become increasingly narrower. Consequently, there is a limit at which a further decrease in $\zeta$ is no longer possible because the deviations in the line wings increase.
Table~\ref{tab-nlte} shows the comparison of the Ba abundances measured in this work and found in the literature obtained either in a 1D or in a 3D NLTE analysis. We find better agreement with the 3D NLTE results of \citep{2020A&A...634A..55G}, while the 1D NLTE gives an abundance that is lower by 0.1 dex on average than ours, possibly due to a slightly higher value of the adopted microturbulence ($\xi= 1$~km s$^{-1}$ in \citealt{2020A&A...634A..55G} versus 0.8~km s$^{-1}$ in our case).

All uncertainties in this analysis lead to a scatter in the abundance values obtained from lines of the given element X. We computed the error [X/H] as the standard deviation over all lines of a given element, and the error in the ratios [X/Fe] as the square root of the sum of the squared errors for [X/H] and [Fe/H].

\section{Results and discussion}

In Table~\ref{tab-a-sun} we report the average values of the solar elemental abundances $A_\odot$  together with data from the literature for comparison.  The obtained abundances of heavy elements agree satisfactorily with the data from other studies (e.g.  \citealt{2021A&A...653A..67B}; \citealt{2015A&A...573A..27G};  \citealt{2019arXiv191200844L}; \citealt{2015MNRAS.446.3651M}; \citealt{2019A&A...624A..78D}).
 \begin{figure}
  \centering
  \includegraphics[width=0.45\textwidth]{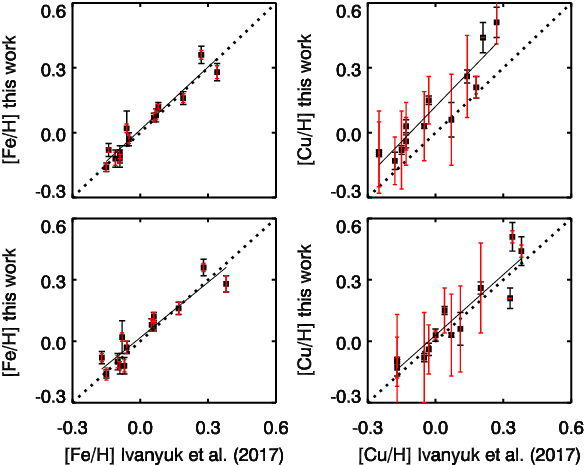}
      \caption{[Fe/H] and [Cu/H] in this analysis using the VALD3 line list and in \citet{2017MNRAS.468.4151I} using the VALD2 line list (top panels).  The same, but using the VALD3 linelist in both analyses (lower panels). The red error bars belong to the  data of \citet{2017MNRAS.468.4151I}, and the black bars belong to the present work. The linear fit to the data is shown by the solid line. The position of the equidistant values is indicated by the dotted line.}
 \label{Fe-Cu}
 \end{figure}

The abundances [X/H] and the relative abundances [X/Fe] for the stars in our sample are given in Tables~\ref{tab-x-h}. We can only compare [Fe/H] and [Cu/H] data from \citet{2017MNRAS.468.4151I} with our results (see Fig.~\ref{Fe-Cu}). In general, the agreement for [Fe/H] is good.  For [Cu/H] there is a systematic difference of about 0.1~dex and large error bars on the [Cu/H]  data of \citet{2017MNRAS.468.4151I} in the majority of the stars.  We used data from VALD3,  whereas \citet{2017MNRAS.468.4151I} used data from VALD2.  When we compared our abundances with those kindly provided by Ivanyuk O.M., obtained using the VALD3 linelist, we found a satisfactory match. This allowed us to conclude that the main reason for the discrepancy is the use of different versions of the VALD database.
  \begin{figure*}
\centering
\includegraphics[width=0.75\textwidth]{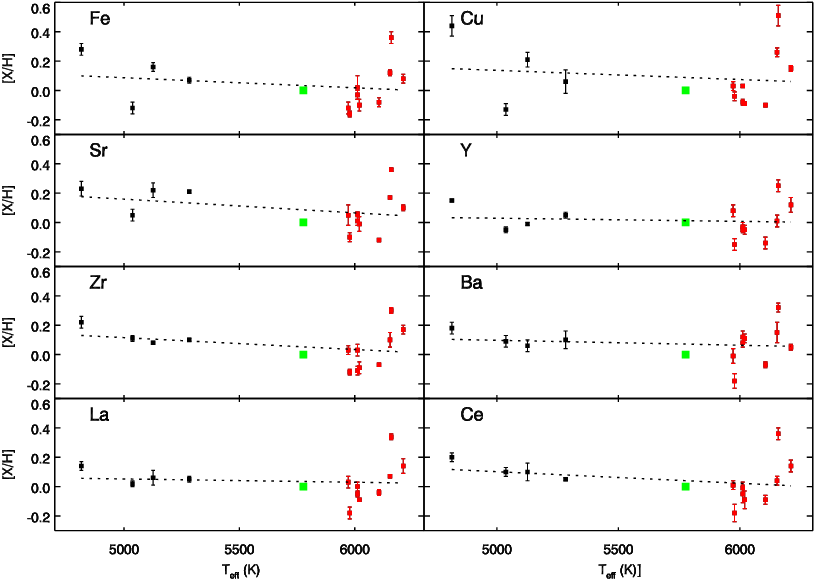}
\caption{[X/H] as a function of   $T_{\rm eff}$. The red squares indicate hot stars with $T_{\rm eff}$ from 5900 to 6200~K, and the black squares indicate cool stars with $T_{\rm eff}$ from 4800 to 5200~K. The Sun is shown with a large green square. The dotted line shows the least-squares fit to the data for all stars.
}
\label{met-tef}
 \end{figure*}
  \begin{figure*}
\centering
\includegraphics[width=0.75\textwidth]{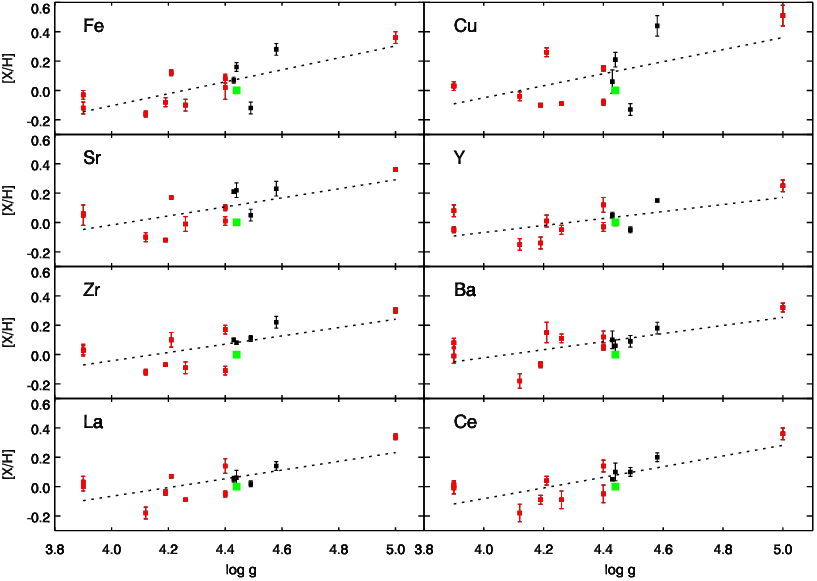}
\caption{[X/H] as a function of  $\log g$. The notations are the same as in Fig.~\ref{met-tef}.
}
\label{met-g}
 \end{figure*}

\begin{figure*}
\centering
\includegraphics[width=0.75\textwidth]{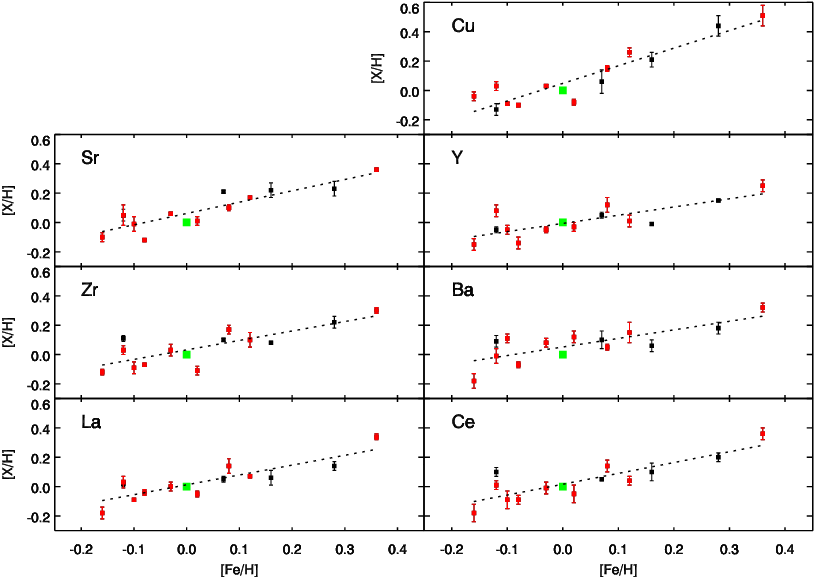}
\caption{[X/H] as a function of metallicity [Fe/H]. The notations are the same as in Fig.~\ref{met-tef}.}
   \label{met-all}
   \end{figure*}

Figures~\ref{met-tef}, \ref{met-g} illustrates the relation between the [X/H] abundance and the stellar parameters $T_{\rm eff}$ and $\log g$.  There is a tendency for the abundance  to increase with  increasing $\log g$  for all elements. We suggest that the two old metal-rich stars HIP 51987 and HD 128358, which may be migrants, cause the [X/H]--$\log g$ trend.
The [X/H]--metallicity  dependence (Fig.~\ref{met-all}) shows little scatter, and it suggests that the abundances of $s$-process elements increase with metallicity.
This agrees with the expected trends in the GCE. In particular, for stars in the thin disk with [X/Fe]~$\approx$~0, there is a consistent trend of increasing [X/H] along with the iron abundance.
Elements heavier than helium are synthesised inside stars and are then ejected by dying stars. The next generation of stars forms from gas clouds in the interstellar medium, which contain heavy elements that were produced and expelled from the previous generation of stars.
   \begin{figure*}[!ht]
\centering
\includegraphics[width=0.75\textwidth]{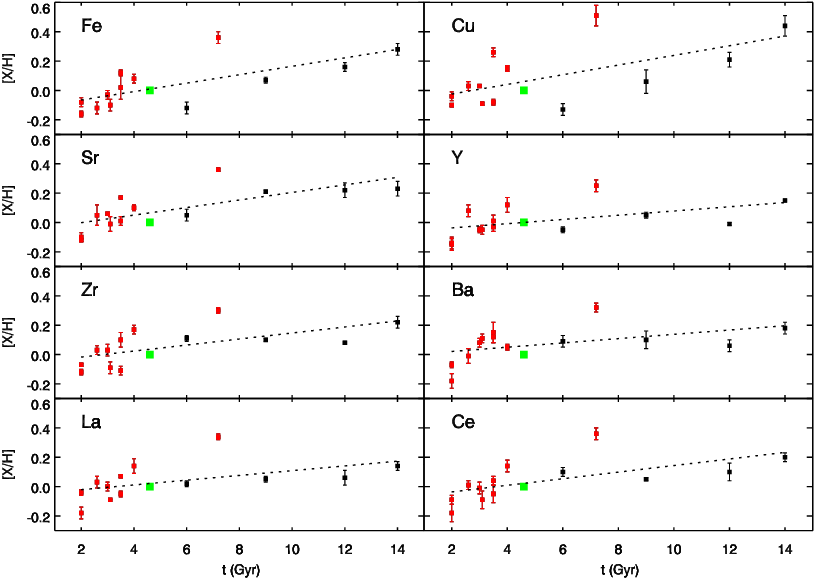}
\caption{[X/H] as a function of  stellar age $t$.  The notations are the same as in Fig.~\ref{met-tef}.}
\label{met-t}
 \end{figure*}
  \begin{figure*}
  \centering
  \includegraphics[width=0.75\textwidth]{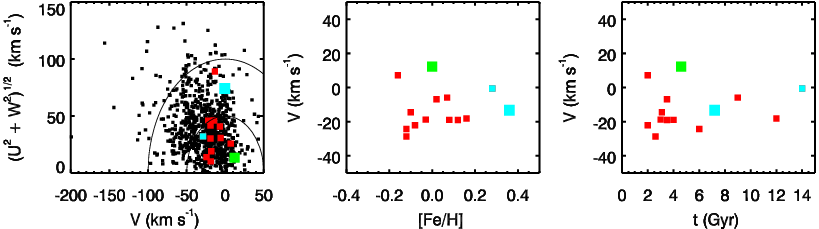}
       \caption{Toomre diagram for the all stars in the CHEPS sample.  The space rotation velocity $V$  vs. [Fe/H] and  vs. stellar age. The solid lines show constant values of the total space velocity, $V_{\rm tot} = (U^2 + V^2 + W^2)^{1/2}$, at 50 and 100 km s$^{-1}$. Our stars are marked in red, HIP 51987 and HD 128358 are shown as large and small blue squares, and the Sun is plotted in green.}
 \label{uvw}
 \end{figure*}

The distribution of [X/H] as a function of stellar age has a large scatter (Fig.~\ref{met-t}). The coverage of [X/H] is far from uniform over the whole age interval because our stellar sample was selected mainly based on metallicity.
However, there are noticeable trends in the increase in the elemental abundances with age. For the $s$-elements, the increase in [X/H] abundances with age is slower than for Fe and Cu.
The literature notes  that there is no clear age--metallicity relation for stars close to the Sun in large samples with a wide range of metallicities (e.g. \citealt{1993A&A...275..101E};  \citealt{2004A&A...418..989N}; \citealt{2014A&A...562A..71B}). A trend of a slowly increasing abundance with decreasing age over a wide range of metallicities from $-1.2$ to 0.5 dex was found in the study of \citet{1993A&A...275..101E}. It was also concluded that 1) most of this scatter is real, not observational; 2) the concept of a well-defined rigid dependence of the metallicity on age is unfounded; and 3) the slope of the [Fe/H] dependence on age during the disc lifetime is very flat, with a large scatter of metallicity at all ages. According to the literature, the obtained [X/H]--age relations show almost flat dependences in the narrow range of $-0.15 \leq$ [Fe/H] $\leq$ 0.4 dex. For example, \citet{2014A&A...562A..71B} have found that in the solar vicinity, stars older than about 8 Gyr show a tendency to decrease in metallicity with age, while younger stars do not show this behaviour. Instead, there is a rather wide spread of ages across the metallicity range ($-0.8$ to +0.4 dex). In other words, there is no clear relation between age and metallicity. \citet{2019A&A...624A..78D} investigated how the abundances of elements with different nucleosynthetic origins evolve with time for a sample of more than 1000 FGK dwarfs from the different populations, including the thin-disk stars. They found a weak age--[Fe/H] correlation, which is thought to be caused by radial migration. Another possible explanation was given by \citet{2020A&A...640A..81N}, who obtained high-precision abundances and determined precise ages from HARPS spectra of 72 nearby solar-type stars. These data suggest that there are two distinct sequences in the age--metallicity diagram. The authors explained this split  as possible evidence of two episodes of accretion of gas onto the Galactic disk with a quenching of star formation in between.

It should be noted that the metallicity-age distribution based on the 107-star sample of  \citet{2017MNRAS.468.4151I} shows a similar tendency to that obtained for our small sample (see Fig.~\ref{par-t-107}). This seems to be due to the peculiarities of the CHEPS project, the aim of which was to search for planets around the most metal-rich solar-type stars. Thin-disk stars were chosen to have spectral types of late-F to mid-K, to have in most cases [Fe/H] $\geq 0.1$ dex, and to be the most inactive. The age of the stars was not considered.  As shown in Fig.~\ref{par-t-107}, a large number of high-metallicity stars are old. In this analysis, we selected stars to cover a wide range of metallicity. As a result, our small sample included  2 metal-rich old stars with ages of $>7$ Gyr. As suggested by \citet{2011A&A...535A..42T} and \citet{2023A&A...669A..96D},  the presence of these stars in regions closer to the Sun may indicate a radial migration from the inner regions of the Galaxy (inner disk and/or bulge). This may explain why the [X/H] abundances obtained for the stars in our sample tend to increase with increasing stellar age.

We checked for any possible star that might migrate from other parts of our Galaxy.
In Fig.~\ref{uvw} we presented the kinematic parameters of our stars $U, V, W$ from \citet{2011A&A...531A...8J}.
Most of the stars are limited by the total spatial velocity of $V_{\rm tot} = 50$~km~s$^{-1}$, with the exception of HIP 51987 and HD 42936. These two stars have a high $V_{\rm tot}$, but it does not exceed 100~km~s$^{-1}$, and   a spatial rotation velocity typical of a thin disk ($V =-0.6$, $-13.3$~km~s$^{-1}$). This means that according to the star kinematics, all our stars belong to the thin disk. On the other hand, HIP 51987 and HD 128356 have  higher metallicities ([Fe/H] = 0.36 and 0.28 dex) and are old ($t$ = 7.2 and 14 Gyr). According to  \citet{2009MNRAS.399.1145S} and \citet{2011A&A...535A..42T} , these two metal-rich stars are old thin-disk stars, and they would also have originated in the inner Galaxy.
  \begin{figure*}
  \centering
  \includegraphics[width=0.75\textwidth]{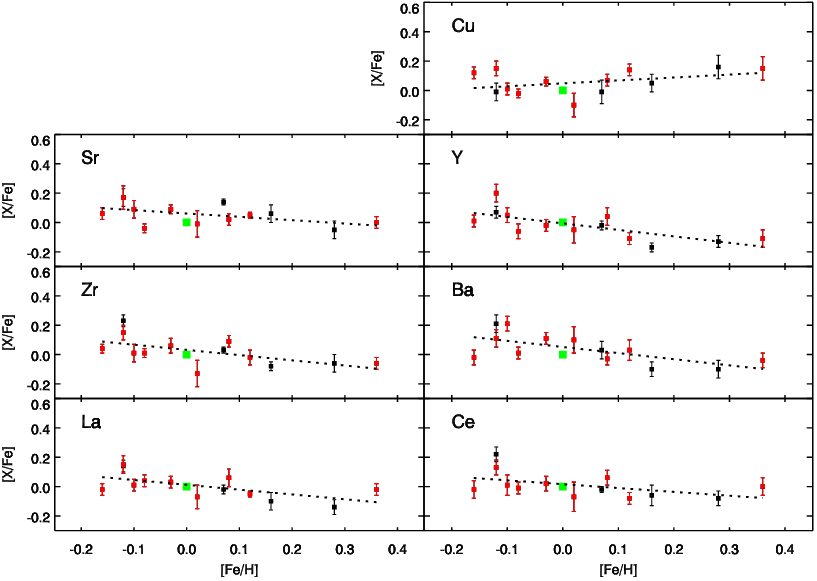}
       \caption{Relative abundances [X/Fe]  vs.  metallicity [Fe/H]. The notations are the same as in Fig.~\ref{met-tef}. }
 \label{met-aFe}
 \end{figure*}
  \begin{figure*}[h!t]
  \centering
  \includegraphics[width=0.75\textwidth]{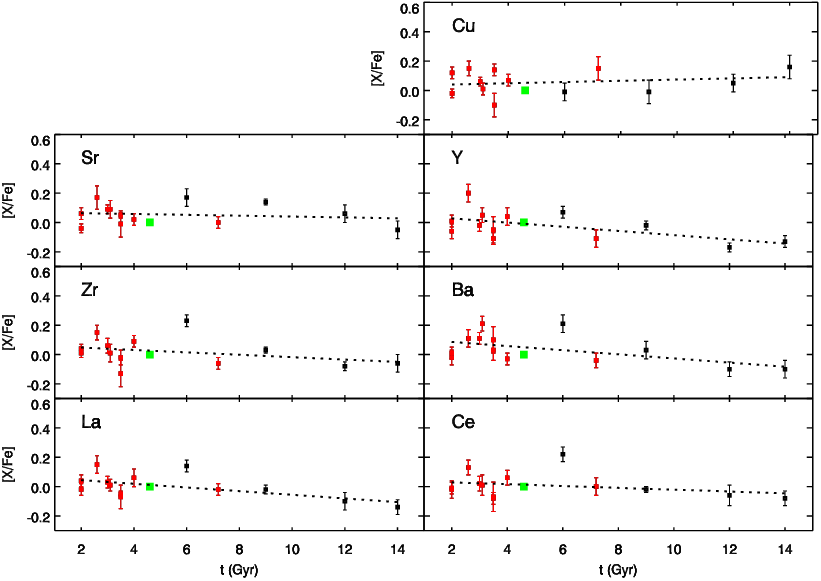}
       \caption{Relative abundances [X/Fe] ratio  vs.  the age $t$. The notations are the same as in Fig.~\ref{met-tef}. }
 \label{met-aFe-t}
 \end{figure*}

In Figs.~\ref{met-aFe} and \ref{met-aFe-t}, the [X/Fe] ratios as a function of metallicity and age are plotted, respectively.  We also quantified these trends for a more robust comparison with the literature and presented the statistical data in Tables~\ref{tab-fit-met}, \ref{tab-fit-t}. The scatter in the [X/Fe]--age relations is smaller than that in the [X/H]--age relations. The trends of increasing [X/Fe] with decreasing metallicity and age are consistent with previous results based on analyses of large samples  of thin-disk stars in the metallicity range of our selected stars. The age and metallicity versus $s$-abundance trends were reported in studies such as \citealt{2014A&A...562A..71B},
\citealt{2015MNRAS.446.3651M}, \citealt{2015A&A...580A..24D},
\citealt{2016ApJ...833..225Z},
\citealt{2018ApJ...865...68B},
\citealt{2018MNRAS.474.2580S}; \citealt{2018A&A...617A.106M},
\citealt{2018AJ....155..111L},
\citealt{2019A&A...624A..78D},
\citealt{2020A&A...639A.127C}, and
\citealt{2021A&A...649A.126T}.  The results of the study of the evolution of the neutron capture elements using detailed GCE  models confirm a decrease in [s/Fe] with increasing [Fe/H] for thin-disk stars (e.g. \citealt{2020MNRAS.492.2828G}; \citealt{2023MNRAS.523.2974M}).  There is an excess of $s$-elements in younger stars. In particular, the [Ba/Fe] excess in our sample is $0.08\pm 0.08$~dex on average for seven stars with an average age of $2.8\pm 0.6$~Gyr. The excess is 0.08 dex higher than the [La/Fe] and [Ce/Fe] ratios and 0.06 dex higher than the average [X/Fe] ratio for all $s$-elements.
It is noteworthy that the [Cu/Fe] ratio has the opposite behaviour. [Cu/Fe] increases with increasing metallicity and age, while the [s/Fe] tends to increase with decreasing metallicity and age.  For [Cu/Fe], the same trends can be found, for example,  in \citet{2011ARep...55..689M} and \citet{2017A&A...606A..94D}.
The similarity of the slopes in the [X/Fe]--metallicity and [X/Fe]--age relations can be explained by the relation between metallicity and age for the stars in our sample (see Fig.~\ref{par-t-107}).
 \begin{figure*}
 \centering
  \includegraphics[width=0.75\textwidth]{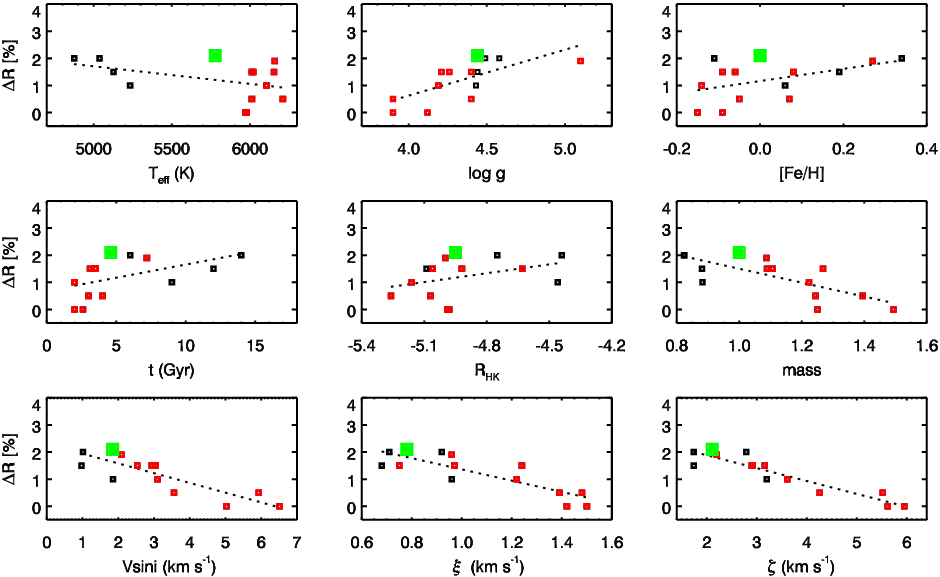}
  \caption {Deviations in the core of the synthesised line from the observed Ba II 649.6 nm line as a function of effective temperature, surface gravity, metallicity, age, chromospheric activity index, rotation velocity, and micro- and macroturbulence. The notations are the same as in Fig.~\ref{met-tef}. }
\label{nlte}
 \end{figure*}

 \begin{figure}
 \centering
  \includegraphics[width=0.4\textwidth]{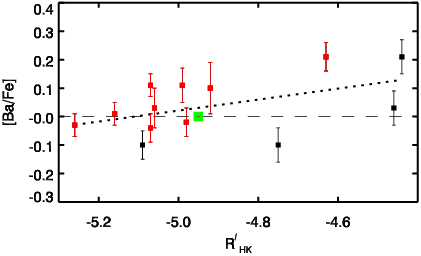}
  \caption {[Ba/Fe] ratio as a function of  the index of the stellar chromospheric activity $R^\prime_{\rm HK}$. The notations are the same as in Fig.~\ref{met-tef}.}
\label{metBa-rhk}
 \end{figure}
 \begin{figure}
 \centering
\includegraphics[width=0.4\textwidth]{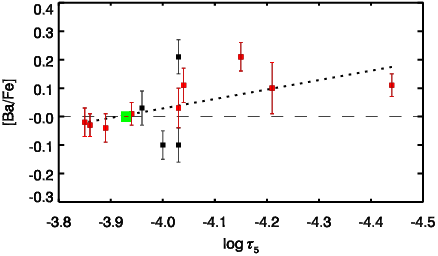}
  \caption {[Ba/Fe]  as a function of the averaged formation depth $\log \tau_5$. The notations are the same as in Fig.~\ref{met-tef}.}
\label{metBa-tau}
 \end{figure}

The obtained trends for stellar abundances can be explained with the help of the GCE model and the periodic table constructed by \citet{2020ApJ...900..179K}.  This table shows that the elements Fe, Cu, Sr, Y, Zr, Ba, La, and Ce were produced by different nucleosynthesis processes in the Galaxy at different times. A large fraction (60\%) of the elements of the iron group is produced at a relatively slow rate by Type Ia supernovae, and the rest is produced by core-collapse supernovae with a mass $> 8-M_{\odot}$. The odd element Cu is mostly produced by core-collapse supernovae and in small amounts by AGB stars.  For these reasons, the abundance [Cu/H] is clearly greater than [Fe/H] (Fig.~\ref{met-all}), and therefore, the [Cu/Fe] ratio increases with metallicity and age (Fig.~\ref{met-aFe}, \ref{met-aFe-t}). The lighter $s$-process elements of the first peak (Sr, Y, and Zr) are produced by low-mass AGB stars and by electron-capture supernovae. The last elements produce 32\% of Sr, 22\% of Y, and 44\% of Zr.  Therefore, we obtained a slightly lower enrichment of stars by the element Y than by the elements Sr and Zr. The main source of the heavier $s$-elements of the second peak (Ba, La, and Ce) is low-mass AGB stars.  According to our findings, the enrichment trends of stars with the elements Ba, La, and Ce generally exhibit similar patterns. However, our data indicate a slightly higher enrichment of the element Ba in younger stars of our sample compared to other second peak $s$-process elements such as La and Ce.  The excess of Ba over La as a function of stellar parameters can be followed in Fig.~\ref{afe-ba-la}. This Ba excess  suggests that an additional factor influences the increase in [Ba/Fe]. It is likely related to the magnetic activity and its impact on the 1D LTE analysis of the Ba lines  \citep{2021A&A...653A..67B}).

The synthesis of Ba II lines has been discussed by many authors because these lines are primarily sensitive to NLTE effects. \citet{1999A&A...343..519M} and \citet{2001A&A...376..232M} showed that the barium line profiles in the solar spectrum cannot be fully reconstructed using 1D NLTE synthesis with reasonable values of $ \log gf$ and height-independent turbulent velocities. The synthetic profiles of the 455.4, 585.3, and 649.6 nm lines were shallower than the observed profiles in the core by 1.7\%, 2.5\%, and 3.8\%, respectively, compared to the observations. Therefore, it has been suggested that introducing height-dependent turbulent velocities into the synthesis is necessary to successfully reconstruct the profiles.  \citet{2020A&A...634A..55G} have carried out a new 3D NLTE synthesis of the Ba lines based on 3D hydrodynamics (HD). They obtained a barium abundance of $A_\odot = 2.27 \pm 0.02 \pm 0.01$ and indicated the systematic and random errors. The inclusion of a more realistic field of non-thermal velocities led to a successful reconstruction of the profiles, except for the 649.6~nm line profile. The line core was still $\approx 2.5$\% shallower than in the observations.

 In our analysis, we also obtained almost the same deviation (2.1\%) in the core of the 649.6~nm line.  The results of the best fit are shown in Figs.~\ref{prof-lines}, \ref{prof-lines1} for the solar barium lines.  Attempts to reconstruct the line core by reducing the macroturbulence were unsuccessful. At $\zeta < 1.3$ km/s, the far wings of the line became narrower than observed. The deviations in the core $\Delta R= (R_{\rm obs} -R_{\rm syn})*100$\% for all stars are shown in Fig.~\ref{nlte}. These deviations were largest in the lines of cooler stars with a higher surface gravity, lower rotational velocities, macro- and microturbulence, and a higher chromospheric activity. The unaccounted-for blend may be one of the reasons for the deviations. In the cooler stars with low velocities, the blend will be stronger, and therefore, the 649.6 nm line core will be deeper and the deviation $\Delta R$ will increase. The second reason may be the sensitivity of the line to the influence of non-thermal velocities. The 649.6 line is  very sensitive to velocities \citep{1993KFNT....9...27S}. In addition, the formation region of the whole line profile extends from the lower photosphere to the lower chromosphere because its effective depth formation in the core is $\log\tau_5 = -4.0$, and in its far wings, it is $\log\tau_5 = -0.7$. In the transition region between the photosphere and the chromosphere, the non-thermal velocities decrease due to the weakening of the granulation motions. Therefore, the lack of a true decrease in the macroturbulence with height will affect the synthesis of these lines most. In other words, if $\zeta$ is constant, it is not possible to recover the core and wings together. In the profile fitting, the real decrease in $\zeta$ with height can be compensated for by the increase in $A_\star$, but it did not affect the [X/H] abundances obtained by the line-by-line approach. The error can only be in the $A_\odot$ and $A_\star$ abundances. We did not exclude the 649.6~nm line, and as a result, we obtained a solar mean barium abundance of $A_\odot =2.30$~dex (Table~\ref{tab-a-sun}), while without this line, $A_\odot =2.27$~dex. This value coincided with the 3D HD NLTE abundance obtained by \citet{2020A&A...634A..55G}.

We considered the dependence of the Ba abundance on the chromospheric activity index of the stars. Fig.~\ref{metBa-rhk} shows that our results confirm the increase in the barium abundance with increasing chromospheric activity. The positive correlation of [Ba/Fe] with the stellar activity index has been reported previously, for example in \citet{2021A&A...653A..67B} and \citet{2017ApJ...845..151R}. This suggests a possible link between chromospheric activity and the formation of the barium lines, which mainly occur in the upper layers of stellar photospheres. If this link exists, then the abundance should depend on the height of the line formation in the atmosphere ($\log  \tau_5$).  In Fig.~\ref{metBa-tau} we plotted the abundances against the height of the line formation in the atmosphere, averaged over the analysed Ba lines. [Ba/Fe] depends on $\log \tau_5$. That is, [Ba/Fe] increases with the height of the line formation.
This dependence can be explained by the high sensitivity of the Ba lines to velocities, whose amplitudes can increase significantly in the upper layers due to chromospheric activity.
A number of other reasons for the increase in [Ba/Fe] has been thoroughly investigated  previously in \citet{2021A&A...653A..67B}. The observed enhancements in Ba might also be a reflection of real variations in abundance. To understand this, it is necessary to use 3D magnetohydrodynamic models that take the influence of chromospheric activity into account, which appears as dark spots and pores, bright dots, plages,  networks, and so on.
In general, the Ba line analyses did not reveal a significant excess of the [Ba/Fe] ratio comparable to that found in young clusters. However, we confirmed that [Ba/Fe] tends to increase with decreasing age in  younger stars of the solar neighbourhood, and we confirmed the increase in the barium abundance with increasing chromospheric activity.

In Figs.~\ref{All-micro}, \ref{All-macro} we presented micro- and macroturbulent velocities as a function of the effective depth of the line core formation. As expected, our results are consistent with data obtained previously with the Fourier method in  \citet{2019KPCB...35..129S}. The microturbulence parameter $\xi$ changes little with height in the atmosphere, while the macroturbulence parameter $\zeta$ tends to decrease with height. The dispersion of the $\zeta$ values is significantly larger than for microturbulence because macroturbulence, as a free parameter, compensates for the effects of NLTE in the lines. Fig.~\ref{All-macro} shows that the values of $\zeta$ deviate from the average in different directions. This is particularly noticeable for the Ba 649.6 nm and Sr 421.5 nm lines. Moreover, these deviations are of opposite sign, corresponding to the individual influence of NLTE effects in these lines. It is known that the solar Ba 649.6 line has a negative NLTE abundance correction (\citealt{2020A&A...634A..55G}; \citealt{1999A&A...343..519M}) and the Sr 421.5 line has a positive correction \citep{2012A&A...546A..90B}. In this analysis, only these two lines show a significant deviation for a macroturbulent velocity. In general, this did not affect the dependence of macroturbulence on height in the stellar atmosphere.

\section{Conclusions}
We have performed a detailed study of the abundances of neutron-capture elements in 13 solar-type F, G, and K stars in the thin disk of the Galaxy.
The stars were selected in a range of metallicities from $-0.15$ to 0.35 dex based on the sample of \citet{2017MNRAS.468.4151I}, with ages from 2 to 14 Gyr.
To obtain elemental abundances, we used high-resolution HARPS spectra and 1D plane-parallel MARCS models of stellar atmospheres. Unlike other analyses, our analysis was based on fitting line profiles with three free parameters: the abundance, and the micro- and macroturbulence. Despite the small number of stars in our sample, important relations were obtained over a sufficient range of stellar ages and were compared with results obtained for large samples of stars.  The dependences of [X/Fe] on metallicity and age agree well with the results obtained by  \citet{2021A&A...653A..67B},  \citet{2020A&A...639A.127C}, \citet{2019A&A...624A..78D}, \citet{2018AJ....155..111L}, \citet{2018A&A...617A.106M}, \citet{2017ApJ...845..151R}, and   \citet{2016ApJ...833..225Z}.  The barium abundance results were analysed in detail. The main findings and conclusions are listed below.
\begin{enumerate}
\item The elemental abundances [X/H] increase with metallicity and age in solar-type stars. The obtained [X/H] increases monotonically with metallicity because these stars with [X/Fe]~$\approx$~0 band belong to the thin disk. The obtained tendency of the abundances [X/H] to increase with age does not confirm the literature data, which indicated a slowly decreasing [X/H] with age or a flat distribution of [X/H] as a function of age in the narrow metallicity range from $-0.15$ to 0.35.
\item The relative abundances [s/Fe] increase with decreasing metallicity and age. In contrast, the [Cu/Fe] ratio increases with increasing metallicity and age. The trends in the obtained dependences agree with the results from a GCE modelling.
\item  The [Ba/Fe] ratio in younger ($<3.5$ Gyr) solar-type stars is 0.06 dex higher than for the other $s$-elements. The [Ba/Fe] ratio increases in stars with higher chromospheric activity. The average excess is found to be $0.15 \pm 0.10$ dex in more active stars. The [Ba/Fe] ratio also increases with increasing line formation height.  This suggests that the influence of active structure features on the line profile and the influence of gradients of non-thermal velocities in the upper layers of the photosphere should be taken into account in line synthesis.
\item The barium 649.6 nm line does not show satisfactory agreement with observations. This is probably due to unknown blends or to the high-velocity sensitivity of the line. We do not recommend to derive the abundances of $A_\odot$ and $A_\star$ from this line.
\item Micro- and macroturbulent velocities were obtained as a byproduct of this synthesis. The macroturbulence decreases with height in the atmosphere, while the microturbulence remains almost unchanged. The macroturbulent velocity, which was used as a free parameter in the LTE synthesis to reconstruct the observed profiles, can correct for NLTE effects in the line core.
\end{enumerate}

\begin{acknowledgements}
We are grateful to the referee for the meticulous examination of our manuscript, the useful and constructive comments, and the valuable recommendations, which have significantly enhanced the quality of our work. We would like to express our sincerest gratitude to Ya. Pavlenko and O. Ivanyuk for their invaluable assistance in the selection of stars and the acquisition of observational data, as well as for their insightful consultations.
            V.D. acknowledges the financial contribution from PRIN-MUR 2022YP5ACE.
\end{acknowledgements}


\bibliographystyle{aa}        
\bibliography{sheminova} 
\newpage
\onecolumn
\begin{appendix}
\section{Additional tables and figures}
\begin{table}[ht]
\centering
 \caption{Intercepts and slopes of linear fits of [X/Fe] ratios  versus [Fe/H].}
\label{tab-fit-met}
{\small
\begin{tabular}{lccc}
\hline\hline
 [X/Fe]   &  Intercept            &     Slope               &   Pearson coefficient \\
 \hline
 {[Cu/Fe]}   &  ~~0.048 $\pm$ 0.021  &      ~~0.200 $\pm$ 0.137  &      ~~0.39  \\
 {[Sr/Fe]}   &  ~~0.061 $\pm$ 0.018  &     $-$0.228 $\pm$ 0.116  &     $-$0.49 \\
 {[Y/Fe]}    & $-$0.006 $\pm$ 0.019  &     $-$0.441 $\pm$ 0.124  &     $-$0.72 \\
 {[Zr/Fe]}   &  ~~0.031 $\pm$ 0.022  &     $-$0.355 $\pm$ 0.142  &     $-$0.58 \\
 {[Ba/Fe]}   &  ~~0.051 $\pm$ 0.021  &     $-$0.417 $\pm$ 0.139  &     $-$0.65 \\
 {[La/Fe]}   &  ~~0.012 $\pm$ 0.018  &     $-$0.328 $\pm$ 0.118  &     $-$0.63 \\
 {[Ce/Fe]}   &  ~~0.016 $\pm$ 0.021  &     $-$0.262 $\pm$ 0.135  &     $-$0.49 \\
 \hline
\end{tabular}
}
\end{table}
\begin{table}[ht]
\centering
 \caption{Intercept and slope of the linear fits of [X/Fe] ratios versus stellar age.}
\label{tab-fit-t}
{\small
\begin{tabular}{lccc}
\hline\hline
 [X/Fe]   &  Intercept            &     Slope               &   Pearson coefficient \\
\hline
{[Cu/Fe]}   &   0.032 $\pm$ 0.039  &      ~~0.004 $\pm$ 0.006  &     ~~0.20 \\
{[Sr/Fe]}   &   0.070 $\pm$ 0.036  &     $-$0.003 $\pm$ 0.005  &    $-$0.15 \\
{[Y/Fe]}    &   0.057 $\pm$ 0.040  &     $-$0.014 $\pm$ 0.006  &    $-$0.56 \\
{[Zr/Fe]}   &   0.064 $\pm$ 0.544  &     $-$0.008 $\pm$ 0.007  &    $-$0.33 \\
{[Ba/Fe]}   &   0.114 $\pm$ 0.042  &     $-$0.014 $\pm$ 0.006  &    $-$0.54 \\
{[La/Fe]}   &   0.069 $\pm$ 0.033  &     $-$0.012 $\pm$ 0.005  &    $-$0.58 \\
{[Ce/Fe]}   &   0.041 $\pm$ 0.040  &     $-$0.006 $\pm$ 0.006  &    $-$0.28 \\
\hline
\end{tabular}
}
\end{table}

 \begin{figure*}[ht]
 \center
  \includegraphics   [width=0.6\textwidth]{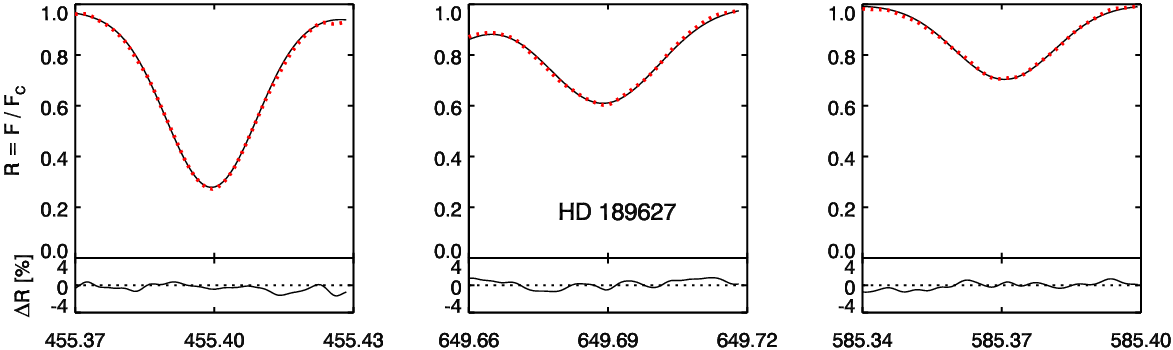}
  \includegraphics   [width=0.6\textwidth]{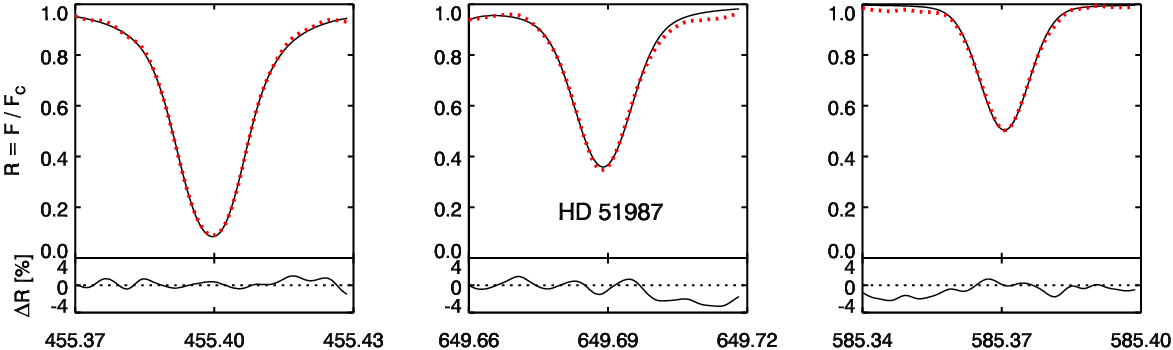}
  \includegraphics   [width=0.6\textwidth]{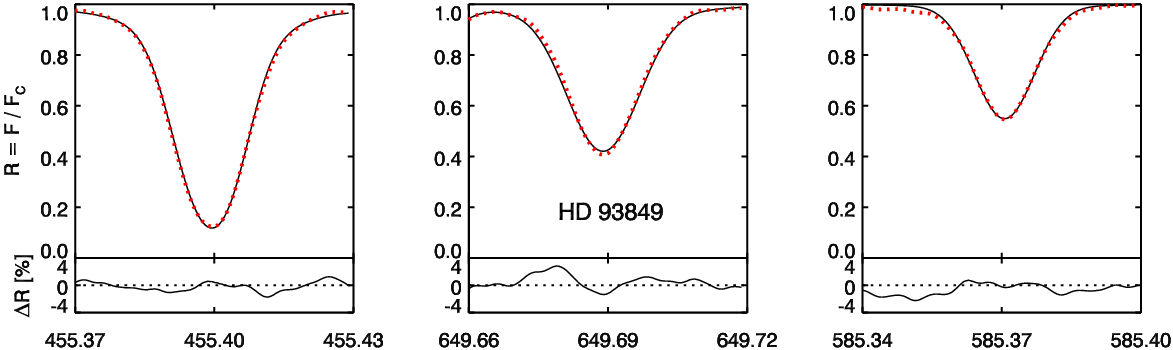}
  \includegraphics   [width=0.6\textwidth]{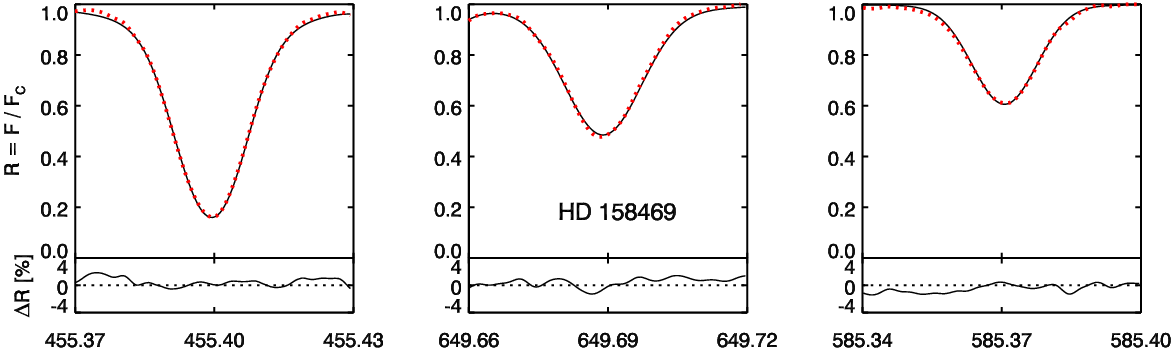}
  \includegraphics   [width=0.6\textwidth]{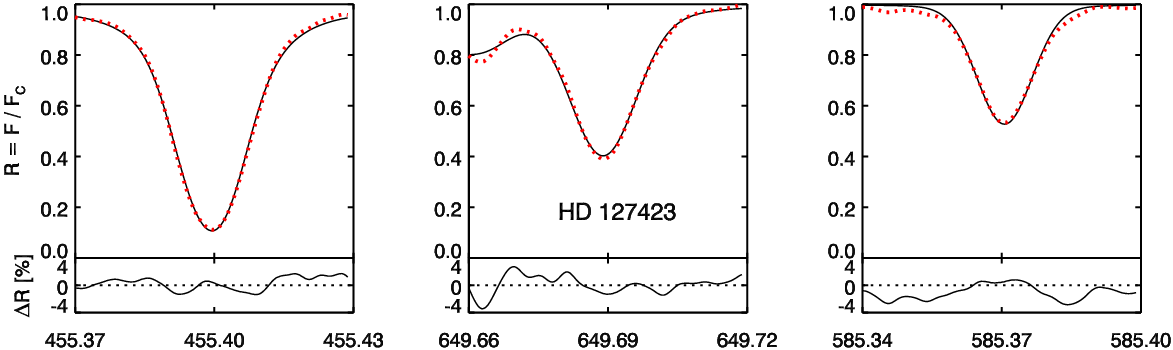}
  \includegraphics   [width=0.6\textwidth]{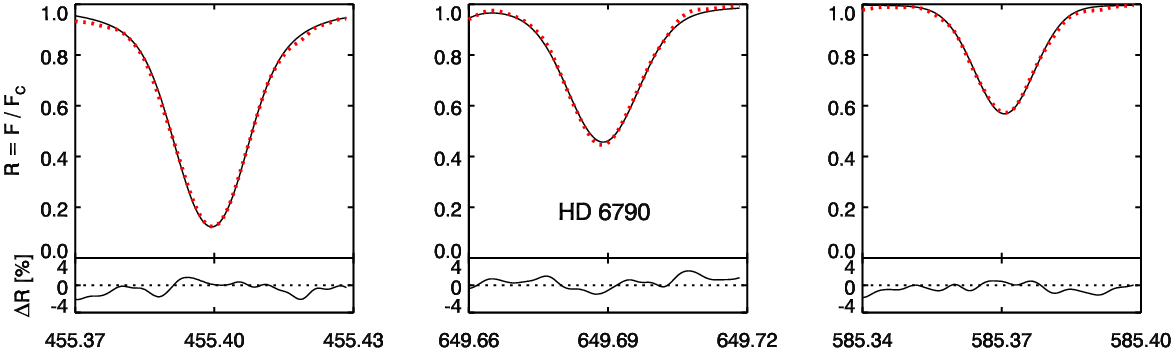}
  \includegraphics   [width=0.6\textwidth]{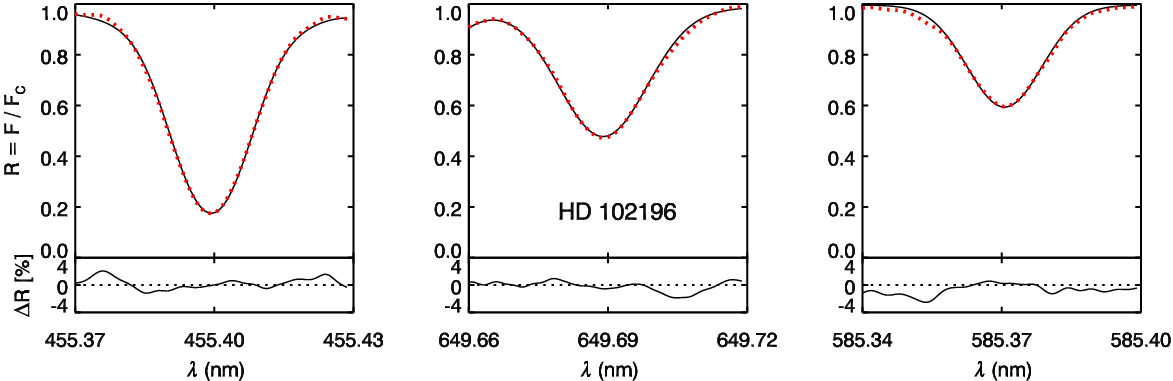}
      \caption {Best fit between synthetic (solid line) and observed (red dashed line) Ba II line profiles in the spectra of the stars with $T_{\rm eff} > 6000$~K.
      The residuals curves ($\Delta R$) show the effect of observational noise in the line profiles. }
\label{prof-lines}
 \end{figure*}
 \begin{figure*}
 \centering
  \includegraphics   [width=0.6\textwidth]{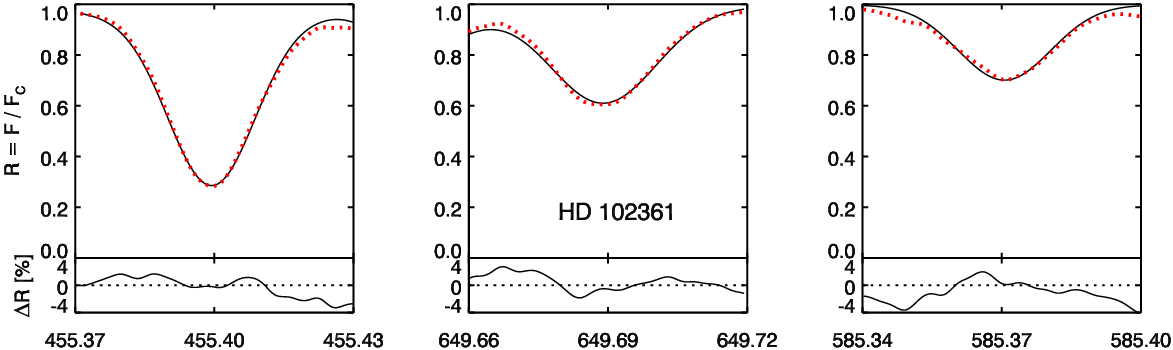}
  \includegraphics   [width=0.6\textwidth]{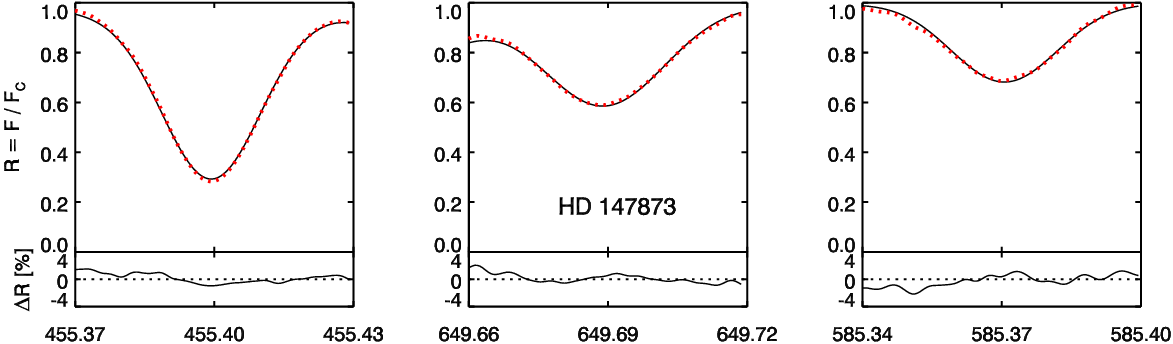}
  \includegraphics   [width=0.6\textwidth]{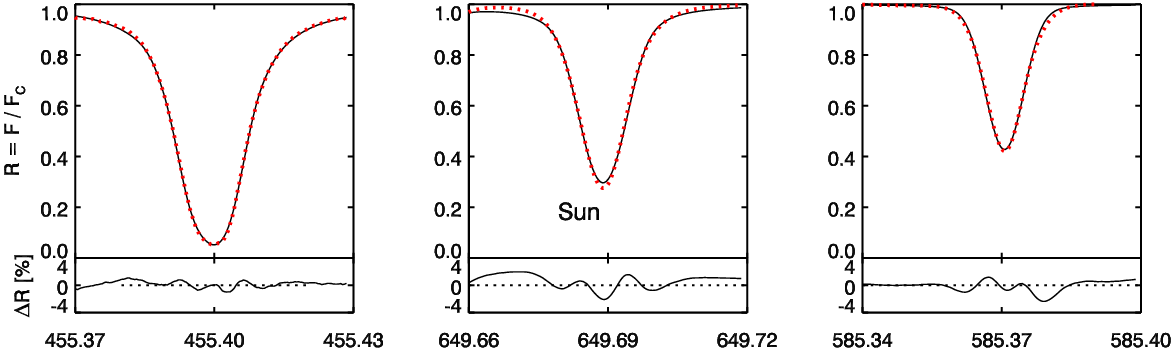}
  \includegraphics   [width=0.6\textwidth]{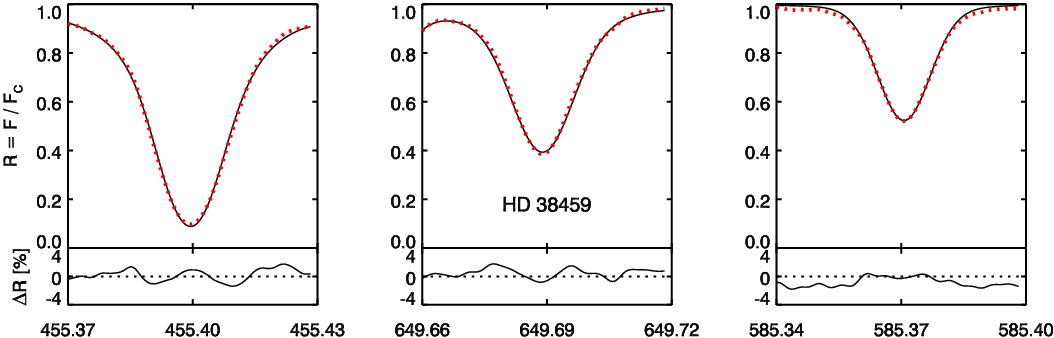}
  \includegraphics   [width=0.6\textwidth]{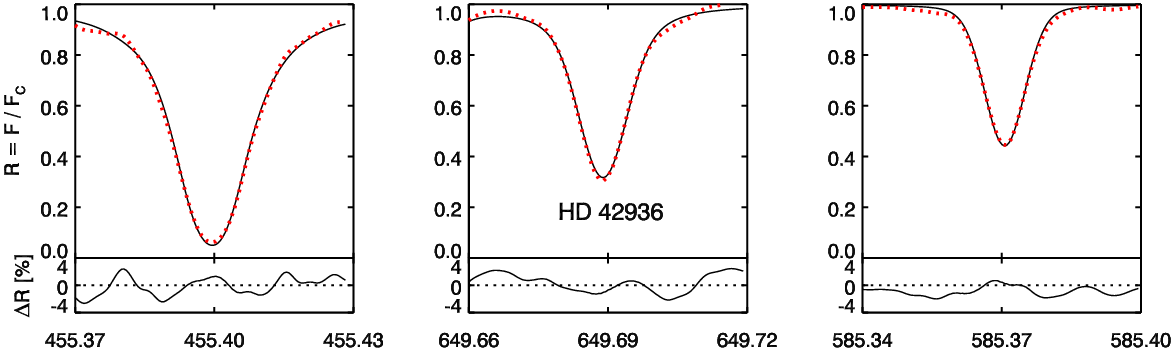}
  \includegraphics   [width=0.6\textwidth]{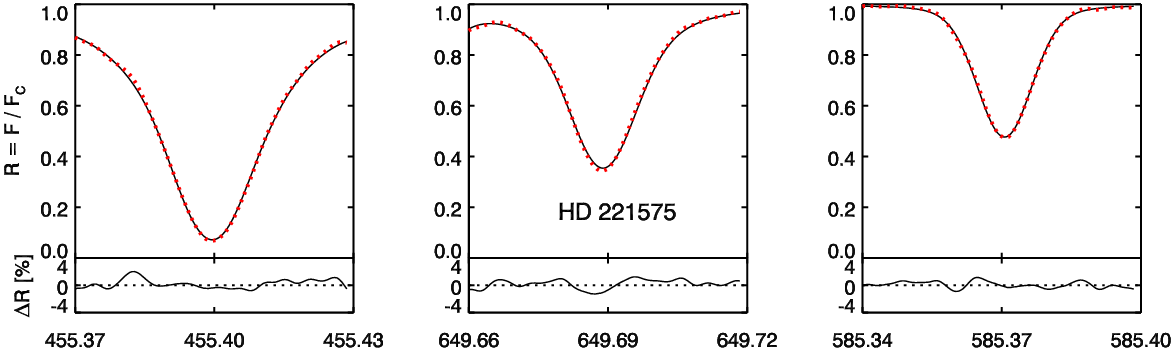}
  \includegraphics   [width=0.6\textwidth]{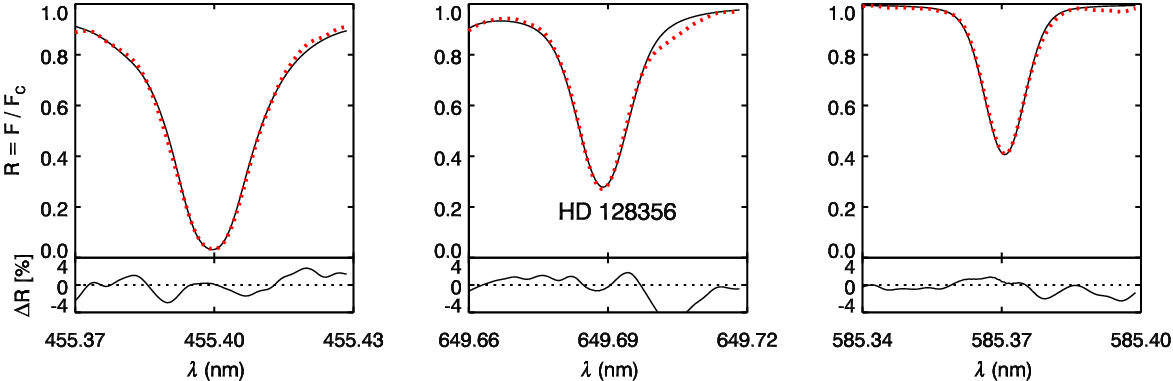}
      \caption {Best fit between synthetic (solid line) and observed (red dashed line) Ba II line profiles in the spectra of the stars with $T_{\rm eff} < 6000$~K. }
\label{prof-lines1}
 \end{figure*}
 \begin{figure*}[ht]
 \centering
  \includegraphics[width=0.75\textwidth]{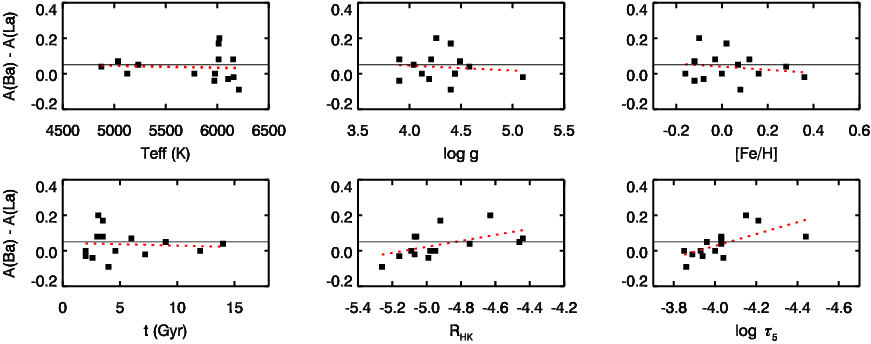}
      \caption {Difference between Ba and La abundances as a function of stellar parameters. The solid line shows that the difference is 0.05 dex. The dotted red line shows the least-squares fit.}
\label{afe-ba-la}
 \end{figure*}

 \begin{figure*}
\centering
  \includegraphics[width=0.85\textwidth]{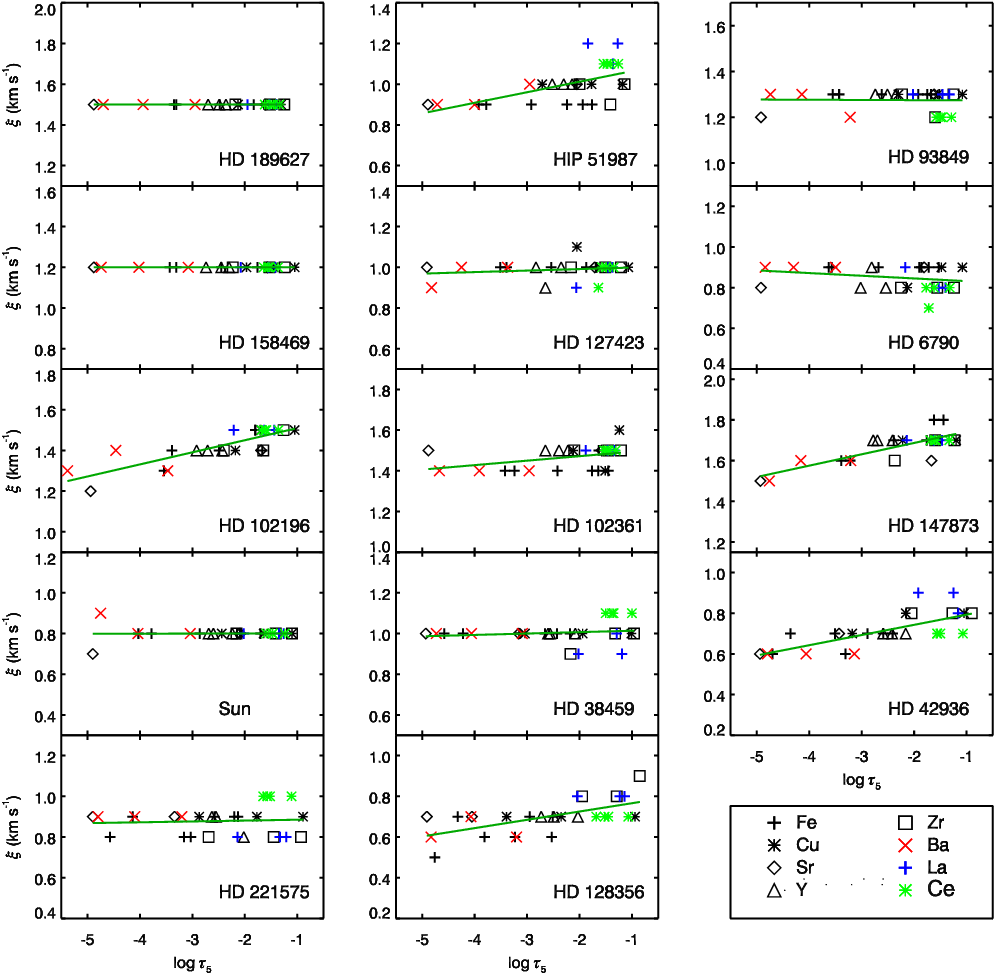}
   \caption {Microturbulent velocities vs. the optical depth of the line core formation. The symbols are notations for the velocities that are derived from the lines of a given element.  The green line shows the least-squares fit to the data.}
\label{All-micro}
 \end{figure*}
 \begin{figure*}
 \centering
  \includegraphics[width=0.85\textwidth]{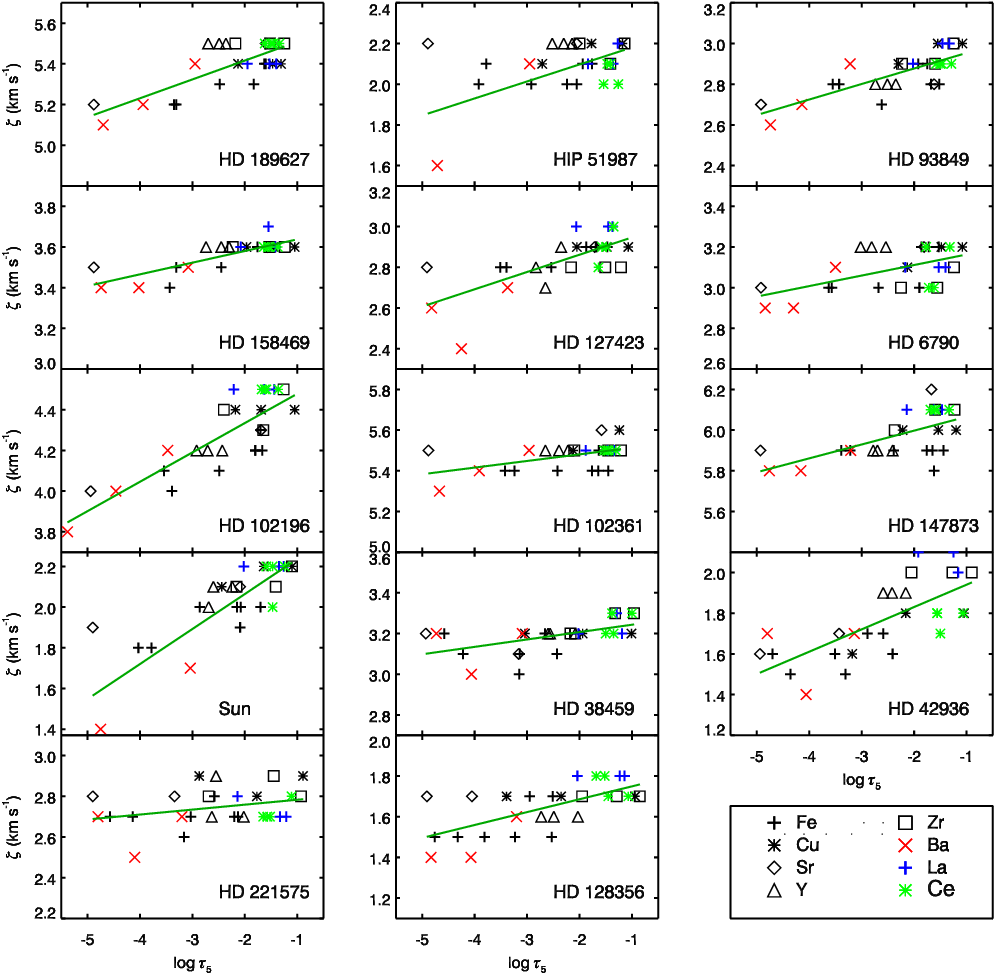}
  \caption {Macroturbulent velocities vs. the optical depth of the line core formation. The notations are the same as in Fig.~\ref{All-micro}}.
\label{All-macro}
 \end{figure*}

\end{appendix}
\end{document}